\documentclass[twocolumn]{aa} %
\usepackage[OT1]{fontenc}
\usepackage{graphicx}
\usepackage{epsfig}
\usepackage{natbib}
\usepackage{natbib}

\def\mathrelfun#1#2{\lower3.6pt\vbox{\baselineskip0pt\lineskip.9pt
  \ialign{$\mathsurround=0pt#1\hfil##\hfil$\crcr#2\crcr\sim\crcr}}}

\def\fun#1#2{\lower3.6pt\vbox{\baselineskip0pt\lineskip.9pt
  \ialign{$\mathsurround=0pt#1\hfil##\hfil$\crcr#2\crcr\sim\crcr}}}

\begin{document}

\title{Testing Gaussianity on Archeops Data}

\author{ 
A.~Curto ~\inst{1} ~\inst{2} \and
J.~Aumont~\inst{3} \and
J.~F.~Mac\'{\i}as--P\'erez ~\inst{3} \and  
E.~Mart\'{\i}nez-Gonz\'alez ~\inst{1} \and
R.B.~Barreiro ~\inst{1} ~\inst{2}\and
D.~Santos~\inst{3} \and
F.~X.~D\'esert~\inst{4} \and
M.~Tristram\inst{3}
}

   \offprints{reprints@archeops.org}

   \institute{
     Instituto de F\'isica de Cantabria, CSIC-Universidad de Cantabria, Avda. de los Castros s/n, 39005 Santander, Spain
    \and
    Dpto. de F\'isica Moderna, Universidad de Cantabria, Avda. los Castros s/n, 39005 Santander, Spain
    \and
    Laboratoire de Physique Subatomique et de Cosmologie, 53 Avenue des Martyrs, 38026 Grenoble Cedex, France
    \and
    Laboratoire d'Astrophysique, Obs. de Grenoble, BP 53, 38041 Grenoble Cedex 9, France
}
   
  \date{\today}
  
  \abstract{} { A Gaussianity analysis using a goodness-of-fit test
     and the Minkowski functionals on the sphere has been
    performed to study the measured Archeops Cosmic Microwave
    Background (CMB) temperature anisotropy data for a 143 GHz
    Archeops bolometer.  We consider large angular scales, greater
    than 1.8 degrees, and a large fraction of the North Galactic
    hemisphere, around 16\%, with a galactic latitude $b > 15$
    degrees.
  }
	   { The considered goodness-of-fit test, first proposed by
    Rayner \& Best (1989), has been applied to the data after a
    signal-to-noise decomposition. The three Minkowski 
      functionals on the sphere have been used to construct a $ \chi^2$ 
      statistic using different thresholds. 
    The first method has been calibrated 
    using simulations of Archeops data containing the CMB signal and
    instrumental noise 
    in order to check its asymptotic convergence.
    Two kind of maps produced with two different
    map-making techniques (coaddition and Mirage) have been analysed.
    }
    {      
      Archeops maps for both Mirage and coaddition map-making, have
    been found to be compatible with Gaussianity.  From these results
    we can exclude a dust and atmospheric contamination larger than
    7.8\% (90\% CL). Also the non-linear coupling
    parameter $f_{nl}$ can be constrained to be $f_{nl} =
    200_{~-800}^{~+1100}$ at the 95\% CL and on angular scales of 1.8
    degrees.
    For comparison, the same method has been applied to data from the
    NASA WMAP satellite in the same region of sky. The 1-year and
    3-year releases have been used. Results are compatible with those
    obtained with Archeops, implying in particular an upper
    limit for $f_{nl}$ on degree angular scales.
    } {}

\keywords{Cosmology
    -- data analysis -- observations -- cosmic microwave background}

\maketitle

\section{Introduction}
According to the inflationary universe theory \citep[see for
example][]{guth_inflation,linde,lyth,liddle}, the primordial density
fluctuations are distributed following very precisely a Gaussian
probability density function (pdf). These fluctuations in the matter
density will produce anisotropies in the temperature of the Cosmic
Microwave Background (CMB) whose pdf is also Gaussian. In this manner,
when the Gaussianity of the CMB radiation is analysed the standard
inflationary theory is tested as well as its alternatives (for example
cosmic strings) which generically predict deviations from it in
different ways.  In addition, the search for non-Gaussianities has
become a powerful tool to detect the presence of residual foregrounds,
secondary anisotropies (such as gravitational lensing,
Sunyaev-Zel'dovich effect) and unidentified systematic errors, which
leave clearly non-Gaussian imprints on the CMB-anisotropies
data. There are many techniques to test Gaussianity, many of them
developed previously as general statistical methods to test the
normality of a data set, and others specifically for the CMB
anisotropies.

Among those methods, we can mention the estimator for non-Gaussianity
based on the CMB bispectrum \citep{ferreira98,magueijo2000},
geometrical estimators on the sphere
\citep{barreiro2001,monteser2005,monteser2006} Minkowski functionals
\citep{meingot,komatsu}, goodness-of-fit tests
\citep{rayner_best_book,aliaga_teoric,barreiro2006}, wavelets
\citep{ferreira97,hobson99,barreiro2000} and steerable filters to
search alignment structures \citep{wiaux}.

Some of them have been applied to the CMB providing different results.
For example WMAP data are compatible with Gaussianity according to the
WMAP team \citep[see][]{komatsu,spergel} whereas others have found
evidences of non-Gausssianities in the same WMAP maps, like
\citet{copi2004,copi2006} (using a technique called multipole vector
framework), \citet{eriksen2004,eriksen2005} (finding asymmetries using
local estimators of the n-point correlations),
\citet{vielva_spot,cruza,cruzb,cruzc} (the Cold Spot detected with
wavelets), \citet{larson} (cold and hot spots different from the ones
expected in Gaussian temperature fluctuations), among others.

In this work the smooth goodness-of-fit test first proposed by
\citet{rayner_best_book} (hereafter R\&BT) will be implemented to
analyse the Gaussianity of the Archeops data. This method has been
already applied successfully to the MAXIMA \citep{cayon} and VSA
experiments \citep{aliaga_vsa,rubino}.  The Archeops data will
be as well analysed with the morphological descriptors known as
Minkowski functionals \citep{schmalzing,meingot}.  The idea is to use
both methods in the Gaussianity analysis for comparison of the
sensitivities of the two techniques and cross-checking of the results
on the amount of dust contamination and the amplitude of the
non-linear coupling parameter.
 
This is the first analysis of Gaussianity of the Archeops experiment
data. We have analysed the data for one of the Archeops bolometer at
143 GHz. This bolometer is the most sensitive and one of the most
relevant for CMB observations. As a complementary analysis, we present
the results of the same goodness-of-fit test applied to WMAP data with
approximately the same mask as the one used for Archeops. The purpose
is to check whether the results are consistent for both data sets.

This paper has the following layout: in Section 2 the R\&BT
applied to signal-to-noise eigenmodes and the Minkowski functionals
are described.  The experiment, main properties of data sets and
masks are summarized in Section 3.  Section 4 is dedicated to the
calibration and checking of both methods with some
``realistic'' CMB anisotropy Gaussian simulations, where we know in
advance the output of the techniques.  Section 5 contains the
Archeops data analysis as well as results. In Section 6 WMAP 1-year
and 3-year data are analysed and compared with Archeops
results. Finally in Section 7 the main conclusions are presented.
\section{Goodness-of-fit tests and Minkowski functionals}
In this section, on the one hand, we will describe briefly
the ``goodness-of-fit technique'' applied to test the Gaussianity of a
set of ``signal-to-noise eigenmodes'' derived from measurements of the
CMB temperature anisotropies. On the other hand, we will
explain the Gaussianity analysis based on the Minkowski functionals.
\subsection {Smooth tests of goodness-of-fit}
Given a set of $n$ random numbers, $\{y_i\}_{i=1}^{i=n}$, it is
sometimes interesting to check whether they behave statistically
according to one specific pdf, $f(y,\theta)$, that is, if the
probability of finding a random number $y$ in an interval between
$y_0$ and $y_0+\Delta y$, with $\Delta y \ge 0$, is given by
$f(y_0,\theta) \Delta y$.  A scalar or vector variable $\theta$ is
introduced, which allows us to move smoothly between different pdf's
in their corresponding space of normalized functions.

This statistical analysis consists in testing the null hypothesis,
$H_0: \{\theta=0\}$ against the alternative hypothesis, $K: \{\theta
\ne 0\}$.

Inside the family of smooth goodness-of-fit tests, we can consider an
{\it order k alternative} pdf $g_k(y, \theta)$, characterized by a pdf
of the form \citep[see][]{rayner_best_book,rayner_best_paper}
\begin{equation}
g_k(y, \theta)=C(\theta)\exp\biggl[\sum_{i=1}^{k}\theta_i h_i(y)\biggr]f(y)
\end{equation}
$\theta$ is a set of $k$ parameters to smoothly cover our space of
pdf's, $f(y)$ is the null hypothesis pdf (e. g. the Gaussian
distribution), $h_i(y)$ form a complete set of orthonormal 
functions\footnote{$\int_{-\infty}^{\infty}h_i(y)f(y)h_j(y)dy=\delta_{ij}$}
on $f(y)$ and $C(\theta)$ is a normalization constant.

The ``score statistic'' is used to evaluate the simple null hypothesis
$H_0$. With this statistic one can estimate the statistical
significance of $\theta$ through the ``Maximum Likelihood Method''.
Following the notation by \citet{aliaga_teoric}, the score statistic
for this goodness-of-fit test is
\begin{equation}
S_k=\sum_{i=1}^k U_i^2 \ \ ,
\end{equation}
and the $U_i^2$ quantities are given by 
\begin{equation}
U_i=\sum_{j=1}^n \frac{h_i(y_j)}{\sqrt n} \label{Ui}
\end{equation}
In the case of a Gaussian pdf, $h_i(x)$ are the ``normalized
Hermite-Chebyshev polynomials''.  If the null hypothesis is satisfied
then the $U_i$ quantities have a statistically normal behaviour and
therefore $U_i^2$ behave like a $\chi_1^2$ distribution
\begin{equation}
f(U_i^2)=\frac{1}{\sqrt{2 \pi U_i^2}}e^{-\frac{-U_i^2}{2}} \label{pdfchi}
\end{equation}
It is possible to write the $U_i^2$ statistical quantities in terms of
the moments of order $k$ derived from the set of $n$ random numbers to
be analysed, $\mu_k=1/n \sum_{j=1}^n y_j^k$, \citep[see for
example][]{aliaga_teoric,aliaga_vsa}.

In this work, the five first statistics $U_i^2$ have been used and can
be related to the $k$-order moments in the following way,
\begin{eqnarray}
\nonumber
U_1^2 & = & n (\mu_1)^2 \\
\nonumber
U_2^2 & = & \frac{n}{2}(\mu_2-1)^2 \\
\nonumber
U_3^2 & = & \frac{n}{6}(\mu_3-3\mu_1)^2 \\ 
\nonumber
U_4^2 & = & \frac{n}{24}(\mu_4-6\mu_2+3)^2 \\ 
U_5^2 & = & \frac{n}{120}(\mu_5-10\mu_3+15\mu_1)^2
\end{eqnarray}
The first few statistics are generally the most sensitive for most of
the applications. In our case higher order $U_i^2$ statistics are
dominated by errors (because of usual propagation of errors) and
therefore are not very useful in practice. This will be described
in detail in section \ref{section4}.
\subsection{Signal-to-noise eigenmode analysis}
At this point, we have described the method that will be used to
analyse a set of $n$ random numbers to test whether their pdf is the
normal distribution or not.

The next step is to compute the set of numbers to be analysed. In our case 
they come from the so called ``signal-to-noise eigenmodes'', firstly
introduced in the CMB field by \citet{stn_eigenmodes}. Our observational data, 
(the fluctuation in the temperature of the incoming blackbody radiation 
measured for each direction $\vec n$ in the sky, $\Delta T ( \vec n)/T$), can
be interpreted as originated from several sources: all the emissions
coming from the sky (CMB signal, Galactic and extragalactic foregrounds and
 atmosphere) and the measured instrumental Gaussian noise 
\citep{archeops_tecnical}.

The total area observed by the experiment is usually divided in equal
area pixels identified by their center direction $\vec n$ and to which
the measurements, $\Delta T ( \vec n)/T$, are assigned. To
obtain the ``signal-to-noise eigenmodes'', we expand the pixel values
of the map, $\Delta T ( \vec n)/T$, into a linear combination in which
the transformed instrumental noise (hereafter the noise) and the
transformed theoretical CMB signal (hereafter the signal) do not have
correlations.

For the signal-to-noise decomposition it is necessary to calculate
signal and noise covariance matrices. The temperature 
covariance between two pixels $i$ and $j$ is given by
\begin{equation}
C_{ij} = \langle \Delta T_i \Delta T_j \rangle -\langle \Delta T_i
\rangle \langle \Delta T_j \rangle \label{corrmatrix} \\ 
\end{equation}
where the brackets $\langle \rangle$ represent the average over several
realizations of temperature anisotropy maps. Thus we can construct the signal
(noise) covariance matrices, $S$ ($N$), averaging on signal 
$\Delta T_s( \vec n)$ (noise $\Delta T_n ( \vec n)$)
realizations. Since the data represent temperature fluctuations around
the mean then it is trivially satisfied that  
$\langle \Delta T_s(\vec n) \rangle =\langle \Delta T_n(\vec n) \rangle = 0$.
Therefore, $C_{ij} = \langle \Delta T_i \Delta T_j \rangle$, the 
correlation matrix.

Once we select a set of $n$ directions in the sky (pixels) and we
construct $S$ and $N$ matrices, which have the same dimension $n$ and
are symmetrical, we can compute the so called ``signal-to-noise
matrix'' $A$
\begin{equation}
A=L_{N}^{-1}SL_{N}^{-t} 
\end{equation}
where $L_N$ is the Cholesky matrix of $N$, defined as $N\equiv L_NL_N^t$. 
$L_N$ can be obtained from the diagonalization of the $N$
matrix. Suppose  $D_N$ is the diagonal  matrix of eigenvalues of $N$,
and $R_N$ a matrix of the eigenvectors of $N$, related by $R_N^tNR_N=D_N$.
Then it is satisfied that $L_N=R_ND_N^{1/2}$, where  
$D_N^{1/2}$ is the square root matrix of $D_N$.

If $\vec d$ is the vector of dimension $n$ representing the data
assigned to the pixels in the sky, the signal-to-noise eigenmodes
can be written as
\begin{equation}
\vec \xi = R_A^tL_N^{-1} \vec d \label{eigenmode}
\end{equation}
where $R_A$ is the matrix of eigenvectors of $A$ and $D_A$ the 
diagonal matrix of eigenvalues of $A$, $R_A^t A R_A = D_A$.

The ${y_i}$ quantities to be analysed with the goodness-of-fit test defined in 
the previous section are
\begin{equation}
y_i=\frac{\xi_i}{\sqrt{1+(D_A)_i}} \label{yi}
\end{equation}
It can be easily demonstrated that if the vector of data $\vec d$
satisfies $\langle \vec d \rangle = 0$ 
then $\langle y_i \rangle = 0$. In the case $\Delta T = \Delta T_s+\Delta T_n$,
from the definition of signal-to-noise 
eigenmodes in (\ref{eigenmode}), the definition of $y_i$ in
(\ref{yi}), and properties of correlation matrices
it follows that $\langle y_i^2 \rangle = 1$.

Supposing that the original map $\vec d$ is multi-normal, then our 
$\{y_i \}$ numbers keep the Gaussian character because both set of
numbers are connected by linear operations. More precisely, they
follow a normal pdf with zero mean and unit variance,
$N(0,1)$. Moreover, for different indexes $i$ and $j$, $y_i$ and $y_j$
are independent.

Finally, for Gaussian data $\vec d$ each $U_i^2$ statistics, defined
in (\ref{Ui}), is distributed as a $\chi_1^2$. The decision to accept
or reject the null hypothesis will be therefore based on this pdf, as
will be seen in sections 5 and 6 when the test is applied to the
Archeops and WMAP data.
\subsection{Minkowski functionals}
Considering the temperature anisotropies of the CMB as a scalar field
on the sphere we can define the set of coordinates $Q_{\nu}$ where
$\Delta T (\vec n) > \nu $ for a given threshold $\nu$, and
its complementary set $V_{\nu}$. As it is stated in
\citet{schmalzing}, any morphological descriptor on the sphere is a
linear combination of 3 Minkowski functionals.  These functionals are:
the area $A(\nu)$ of the excursion set $Q_{\nu}$, the contour
length $C(\nu)$ of the excursion set $Q_{\nu}$, and the genus
$G(\nu)$ (defined as the number of hot spots above $\nu$ minus
the number of cold spots below that threshold).

For a Gaussian random field the mean values of these
functionals are
\begin{eqnarray}
\nonumber
\langle A(\nu) \rangle &   = &  \frac{1}{2} \left(1  - \frac{2}{\sqrt{\pi}}\int_{0}^{\nu/2}exp(-t^2)dt\right) \\
\label{area}
\nonumber
\langle C(\nu) \rangle & = &  \frac{\sqrt{\tau}}{8}exp(-\frac{\nu^2}{2})  \\
\label{contour}
\nonumber
\langle G(\nu) \rangle & =  & \frac{\tau}{(2\pi)^{3/2}} ~ \nu ~ exp( -\frac{\nu^2}{2})\\
\label{genus}
\end{eqnarray}
where $ \tau$ is a parameter related with the coherence angle
\citep{barreiro2001,schmalzing}.

The Gaussianity test with the Minkowski functionals is performed
through a $ \chi^2$ test as described for example in
\citet{komatsu,spergel}. Considering $ n_{th}$ possible thresholds $
\nu$ we can define a $ 3n_{th}$ vector $ \vec v = (A(\nu), C(\nu),
G(\nu))$. The $ \chi^2$ statistic is then defined
\begin{equation}
 \chi^2 = \sum_{i,j}(\vec v(i) - \langle \vec v(i) \rangle)C^{-1}_{ij}(\vec v(j) - \langle \vec v(j) \rangle)
\end{equation}
where $ \langle \vec v(j) \rangle$ is the expected value of
$ v(j)$ and $ C$ is the corresponding covariance matrix for all possible
thresholds and functionals.
\section{Archeops data sets}
\subsection{The Archeops experiment}
Archeops\footnote{\tt http://www.archeops.org} is a balloon borne
experiment dedicated to measure the CMB temperature anisotropies from
large to small angular scales \citep{archpaper,tristram_cl}.  It has
given the first link in the $C_\ell$ determination between the COBE
large angular scales data \citep{cobe} to the first acoustic peak as
measured by BOOMERanG and MAXIMA \citep{boomerang,maxima}.  Archeops
was also designed as a test bed for the forthcoming Planck High
Frequency Instrument (HFI), \citep{lamarre}.  Therefore, Archeops
shared with Planck the same technological design: a Gregorian off-axis
telescope with a 1.5 m primary mirror, bolometers operating at 143,
217, 353 and 545 GHz cooled down at 100 mK by a $^3$He/$^4$He dilution
designed to work at zero gravity and a similar scanning
strategy. Archeops was launched on February 7$^{\rm th}$, 2002, from
the CNES/Swedish facility of Esrange, near Kiruna (Sweden). 12
hours of high quality night data were gathered. This data corresponds
to a coverage of approximately 30\% of the sky, including the Galactic
plane. More details about the instrument and the flight performance
can be found in \citet{archpaper_cospar,archeops_tecnical}. From its
four frequency bands the two lowest (143 and 217 GHz) were dedicated
to the observation of the CMB and the others (353 and 545 GHz) to the
monitoring and calibration of both atmospheric and Galactic emissions.

In the following, we focus on the analysis of the most sensitive
143 GHz Archeops bolometer that also presents the lowest level of
contamination by systematic effects.

Although the Archeops resolution is typically of 10 arcmin, for this
analysis we are interested in the Gaussianity of the large angular
scale anisotropies. Therefore, we decided to use low resolution maps
at HEALPix \citep{healpix} $N_{\rm side}=32$ to consider scales above
1.8 degrees. 
\subsection{Data processing}
We describe here briefly the way that Archeops data were
processed. For a more detailed description see
\cite{archeops_tecnical}.

In the Time Ordered Information (TOI) corrupted data are flagged
(representing less than 1.5\% of the whole data set). Low frequency
drifts, correlated to house-keeping data are removed using the latter
as templates. A high frequency decorrelation is also performed to
remove some bursts of non-stationary high-frequency noise. Corrected
timelines are then deconvolved from the bolometer time constant and
the flagged corrupted data are replaced by a realization of
noise. Finally, low time frequency atmospheric residuals are
subtracted using a destriping procedure which slightly filters out the
sky signal to a maximum of 5\%.

Archeops cleaned TOIs at 143 GHz are contaminated by atmospheric and
Galactic dust residuals, even at intermediate Galactic
latitudes. Atmospheric residuals contributes mainly at frequencies
lower than 2 Hz in the timeline and follows approximatively a $\nu^2$
law in antenna temperature. Galactic dust presents a grey body
spectrum at about 17 K and with an emissivity of about $\nu^2$. To
suppress both residual dust and atmospheric signals, data are
decorrelated using a linear combination of the high frequency
photometric pixels (353 and 545 GHz) and of synthetic dust
timelines.

We have used in this work two kind of map-making for the TOIs of
Archeops data and of the simulations. The first one is an optimal
map-making procedure called Mirage \citep{mirage}. Mirage is based on
a two-phase iterative algorithm, involving optimal map-making together
with low frequency drift removal and Butterworth high-pass
filtering. A conjugate gradient method is used for resolving the
linear system. The second is a procedure that performs
coaddition. This means that all the TOI points corresponding to a
given pixel are averaged.

\begin{figure}[!t]
\center
\epsfig{file=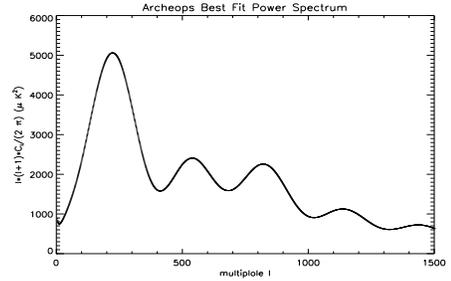,height=3.8cm,width=6.0cm}
\caption{Archeops Best Fit Power Spectrum used to simulate the Archeops 
CMB signal.
}
\label{archeops_cl}
\end{figure}

To produce a CMB simulation, a random CMB map with the power spectrum
of the Archeops model (see \citet{archpaper_cospar} and figure
\ref{archeops_cl}) is generated and from this map an Archeops TOI is
produced. This TOI is treated with the two map-making methods
described above to produce a map. To perform a noise simulation we
produce a Gaussian constrained realization of the Archeops noise power
spectrum in the time domain. The TOI produced this way is then
projected into a map using the above map-making techniques.

\label{simus}

\begin{figure}[!t]
\center
\epsfig{file=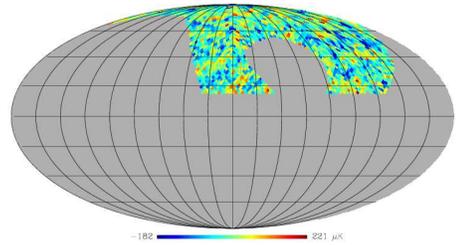,height=3.8cm,width=6.0cm}
\caption{Mirage Archeops data from the best bolometer at 143 GHz 
presented  at HEALPix resolution $N_{side}=32$, 
($\approx 1.8$ degrees). This map is centered on Galactic longitude 
$l$ = 180 degrees. Galactic and South Equator pixels have been masked.
Grid lines are spaced by 20 degrees.
}
\label{skycoverage}
\end{figure}

The analysis has been performed on a fraction of the Archeops observed
region masking out pixels with Galactic latitude below 15 degrees,
$|b|<15^o$. The southern sky data were not included in the
analysis as they are more contaminated by systematics in the form of
residual stripes coming from the Fourier filtering and destriping of
the data in the time domain \citep{archeops_tecnical} that produces
ringing around the Galactic plane.  In the case of the CMB power
spectrum analysis presented in \citet{tristram_cl} this southern sky
region was used as increased significantly the signal to noise ratio
at small angular scales which are not affected by this systematic
effect.  This is not the case for the analysis presented in this paper
where we are more interested in large angular scales where this
systematic becomes important.
In figure \ref{skycoverage} we plot the region of data considered for
the analysis. These data correspond to 1995 pixels (16\%
of the sky) from a total of 12288 pixels for a complete map at this
resolution.
\def\lea{\mathrel{\raise .4ex\hbox{\rlap{$<$}\lower 1.2ex\hbox{$\sim$}}}}
\def\gea{\mathrel{\raise .4ex\hbox{\rlap{$>$}\lower 1.2ex\hbox{$\sim$}}}}
\let\lsim=\lea
\let\gsim=\gea
\section{Calibrating the method: analysis on Gaussian simulations}
\label{section4}
\begin{figure}[t]
\center
\epsfig{file=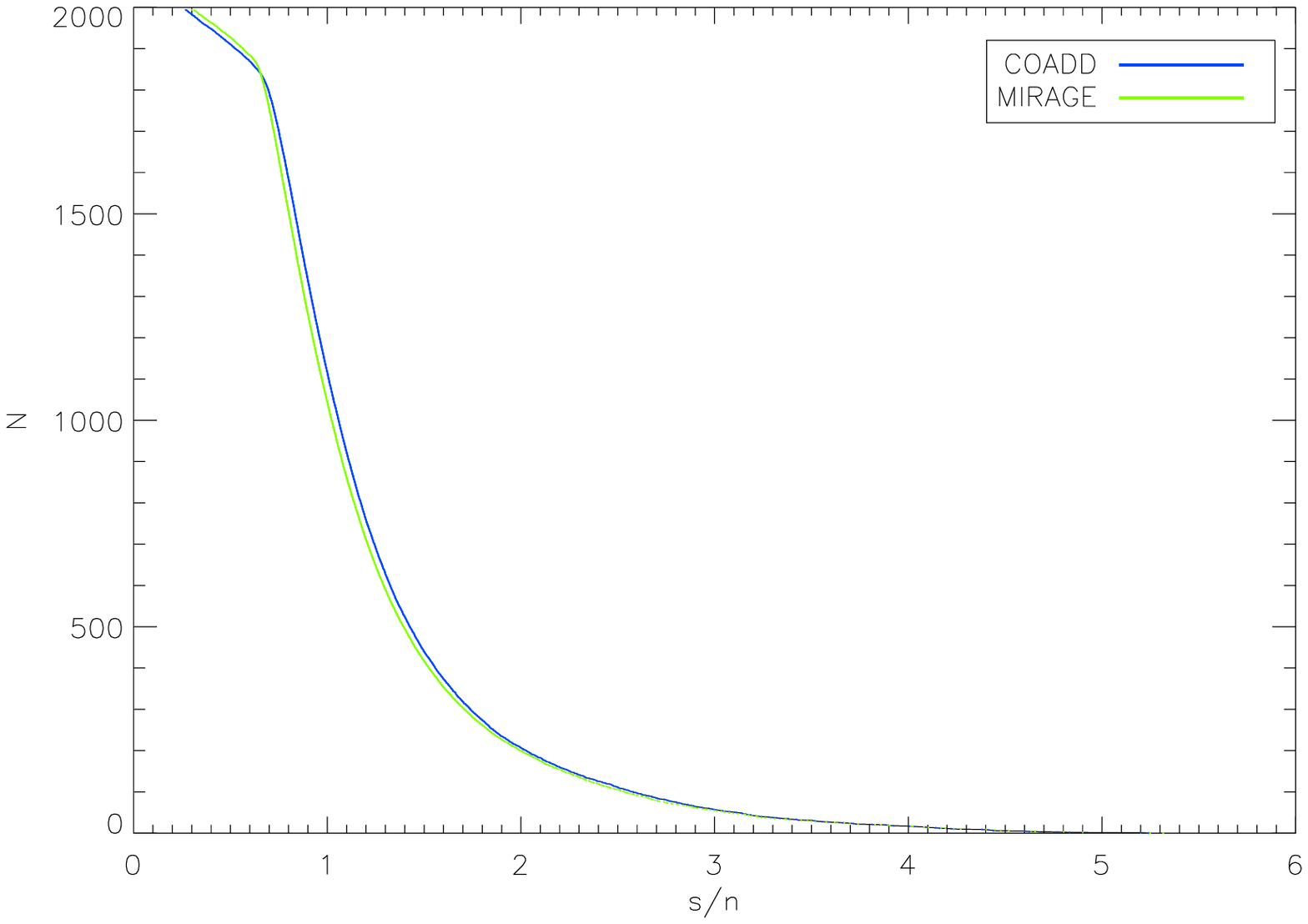,height=3.8cm,width=6.0cm}
\caption{Number of normalized signal-to-noise eigenmodes $y_i$ for which their 
  associated $A$ matrix eigenvalues, $(D_A)_i$, satisfy 
  $(D_A)_i \ge (s/n)_c^2$.
}
\label{stncuts_vs_n}
\end{figure}
\begin{figure*}[t]
\center 
\epsfig{file=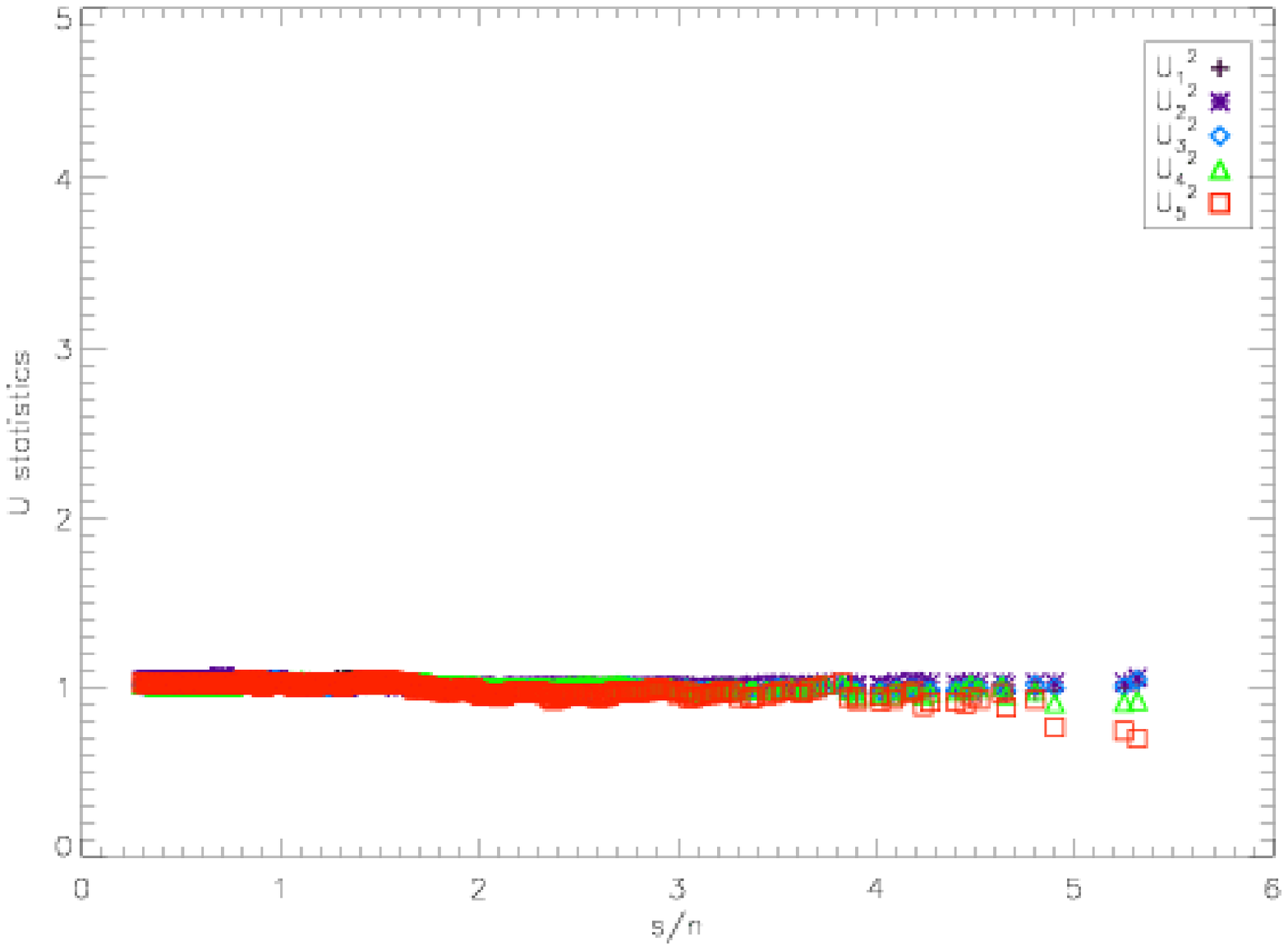,height=3.8cm,width=6cm}
\epsfig{file=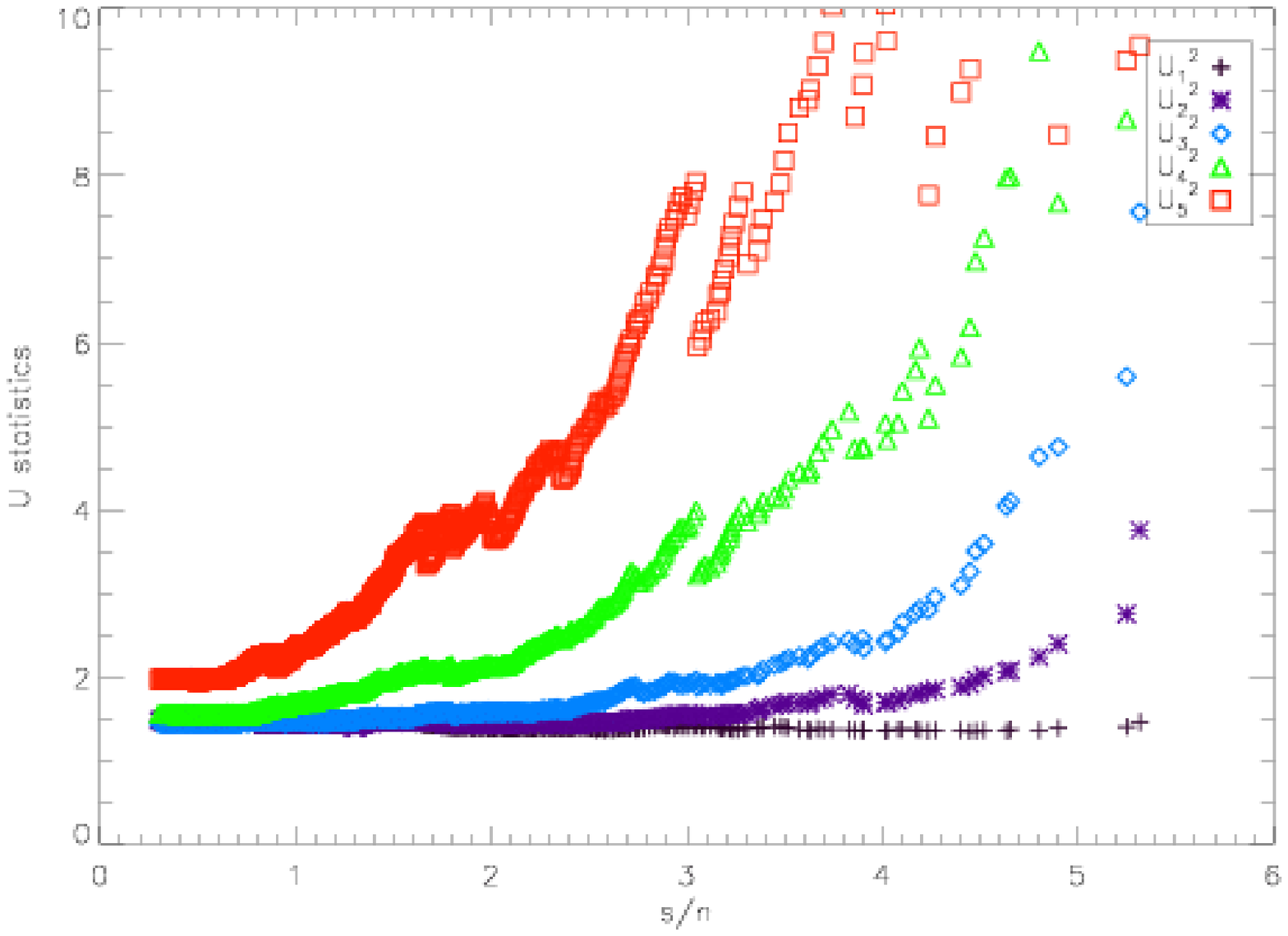,height=3.8cm,width=6cm}
\caption{{\it \textbf{From left to right,}} mean and dispersion of
$U_i^2$ statistics (where $i$ goes from 1 to 5) for different
signal-to-noise cuts, corresponding to $10^4$ signal plus noise Mirage
simulations. }
\label{mu_stat_simu_mirage}
\end{figure*}
\begin{figure*}[t]
\center
\epsfig{file=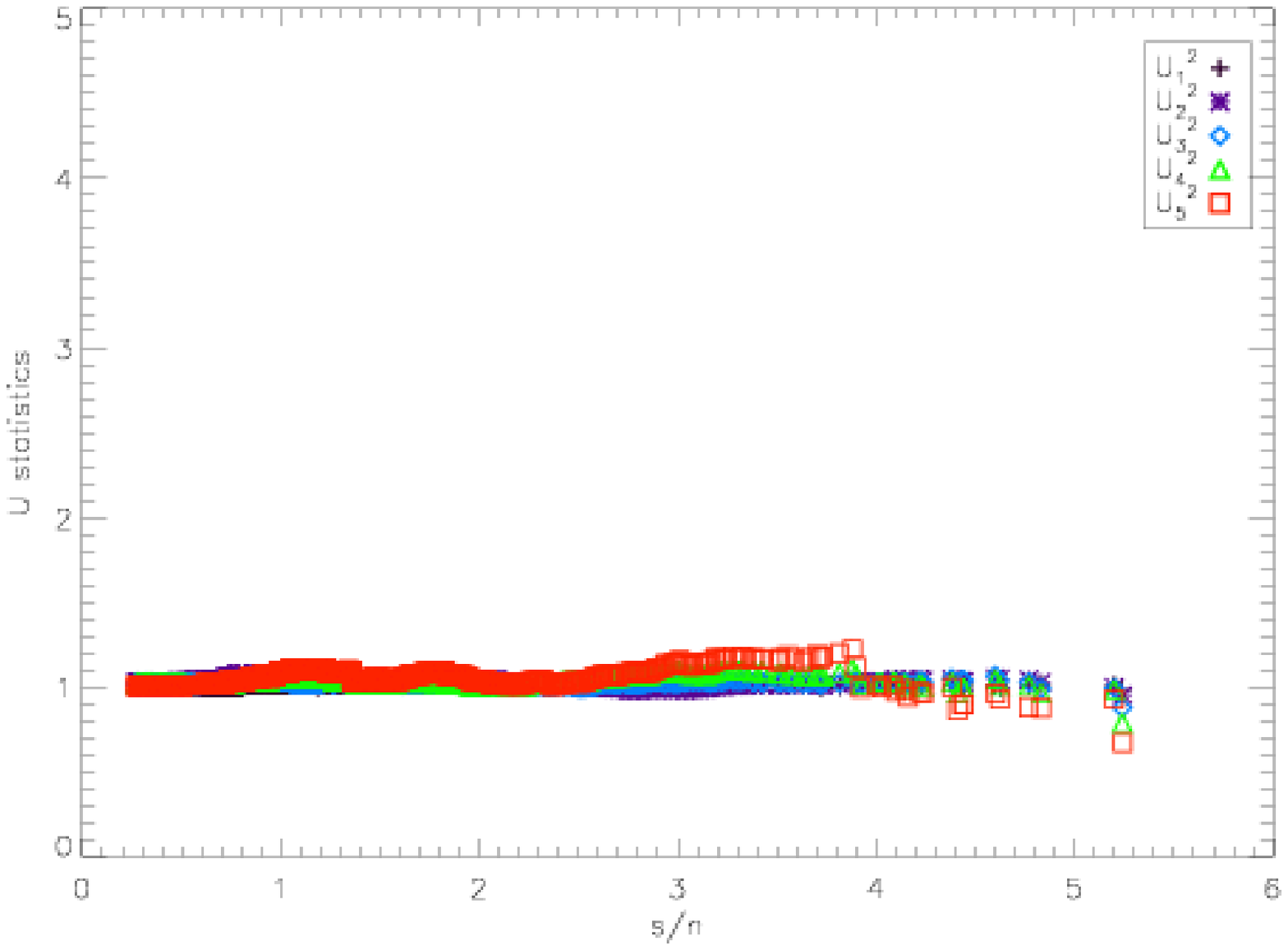,height=3.8cm,width=6cm}
\epsfig{file=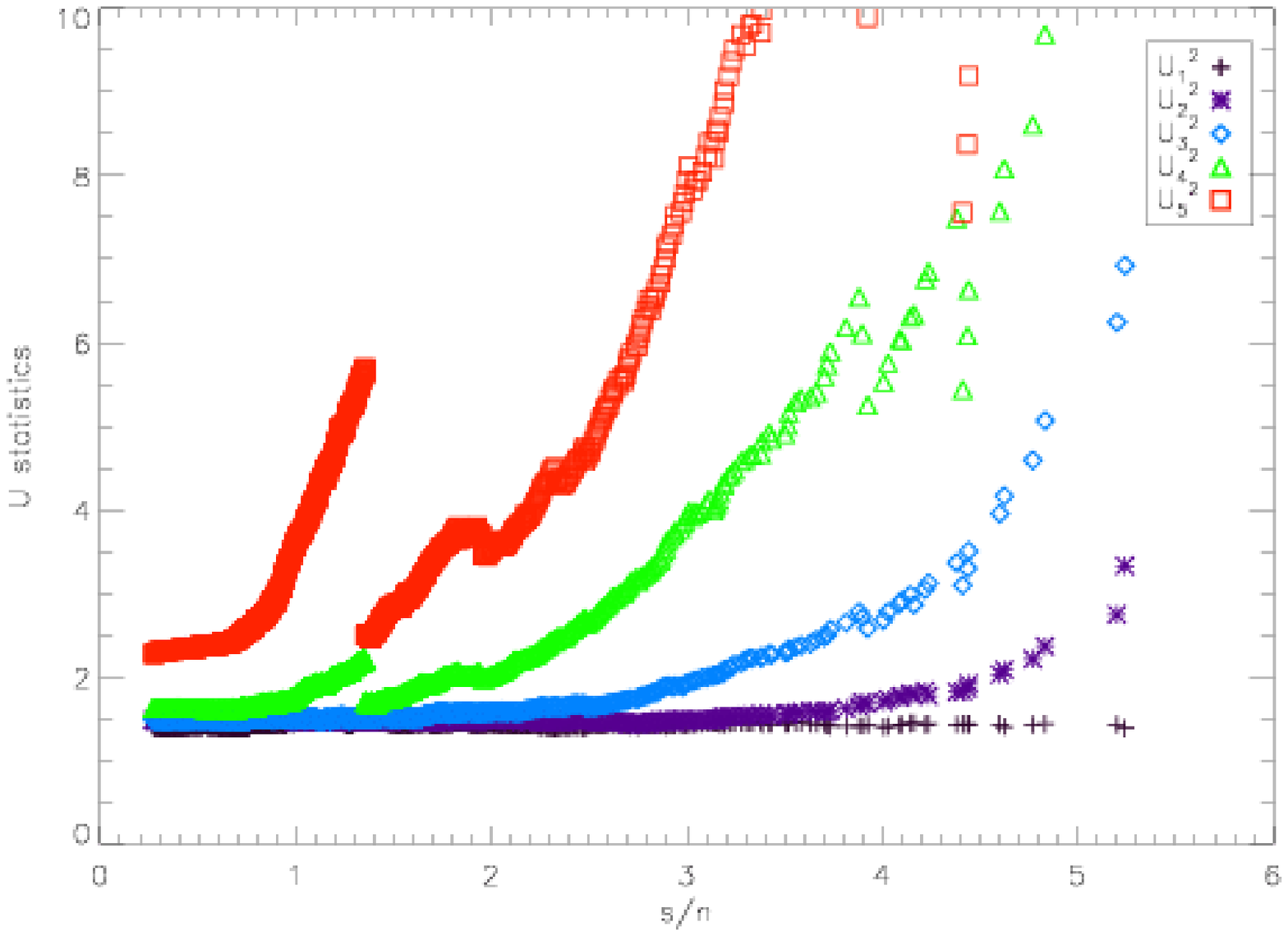,height=3.8cm,width=6cm}
\caption{{\it \textbf{From left to right,}} mean and dispersion of
$U_i^2$ statistics (where $i$ goes from 1 to 5) for different
signal-to-noise cuts, corresponding to $10^4$ signal plus noise
coaddition simulations.}
\label{mu_stat_simu_coadd}
\end{figure*}
\begin{figure*}[tb]
 \begin{center}
  \includegraphics[angle=0,height=4.0cm,width=4.0cm]{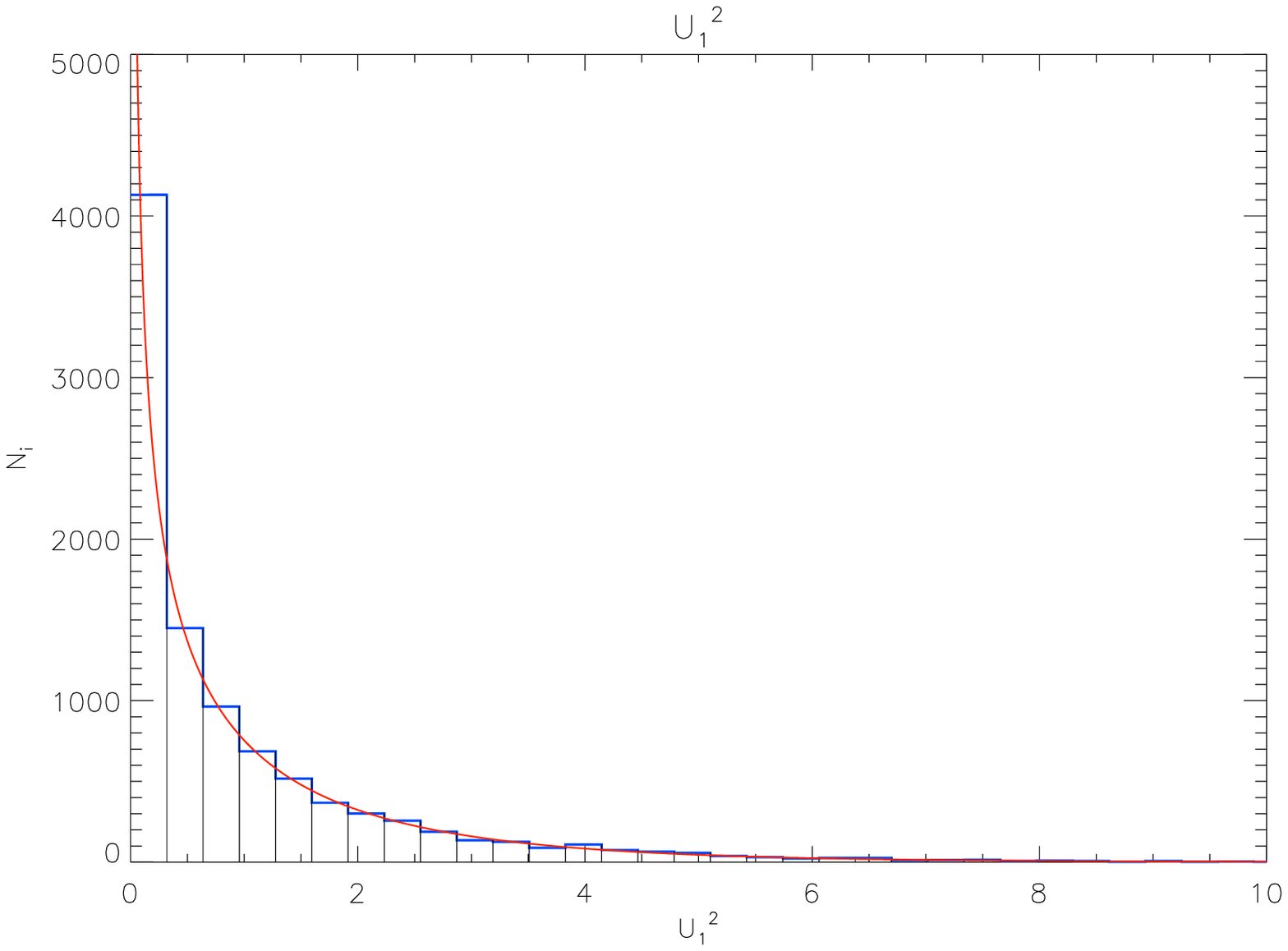}
  \includegraphics[angle=0,height=4.0cm,width=4.0cm]{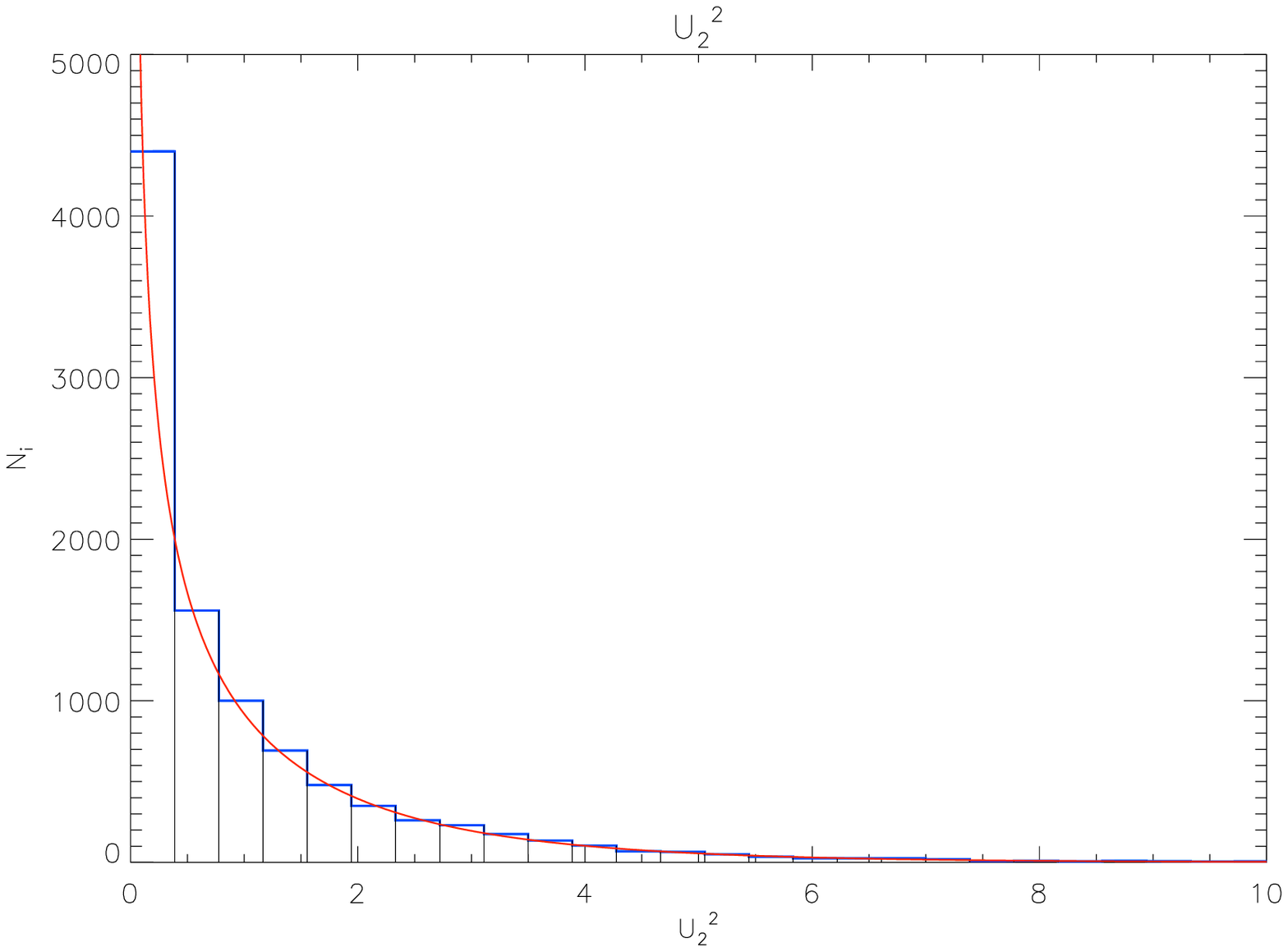}
  \includegraphics[angle=0,height=4.0cm,width=4.0cm]{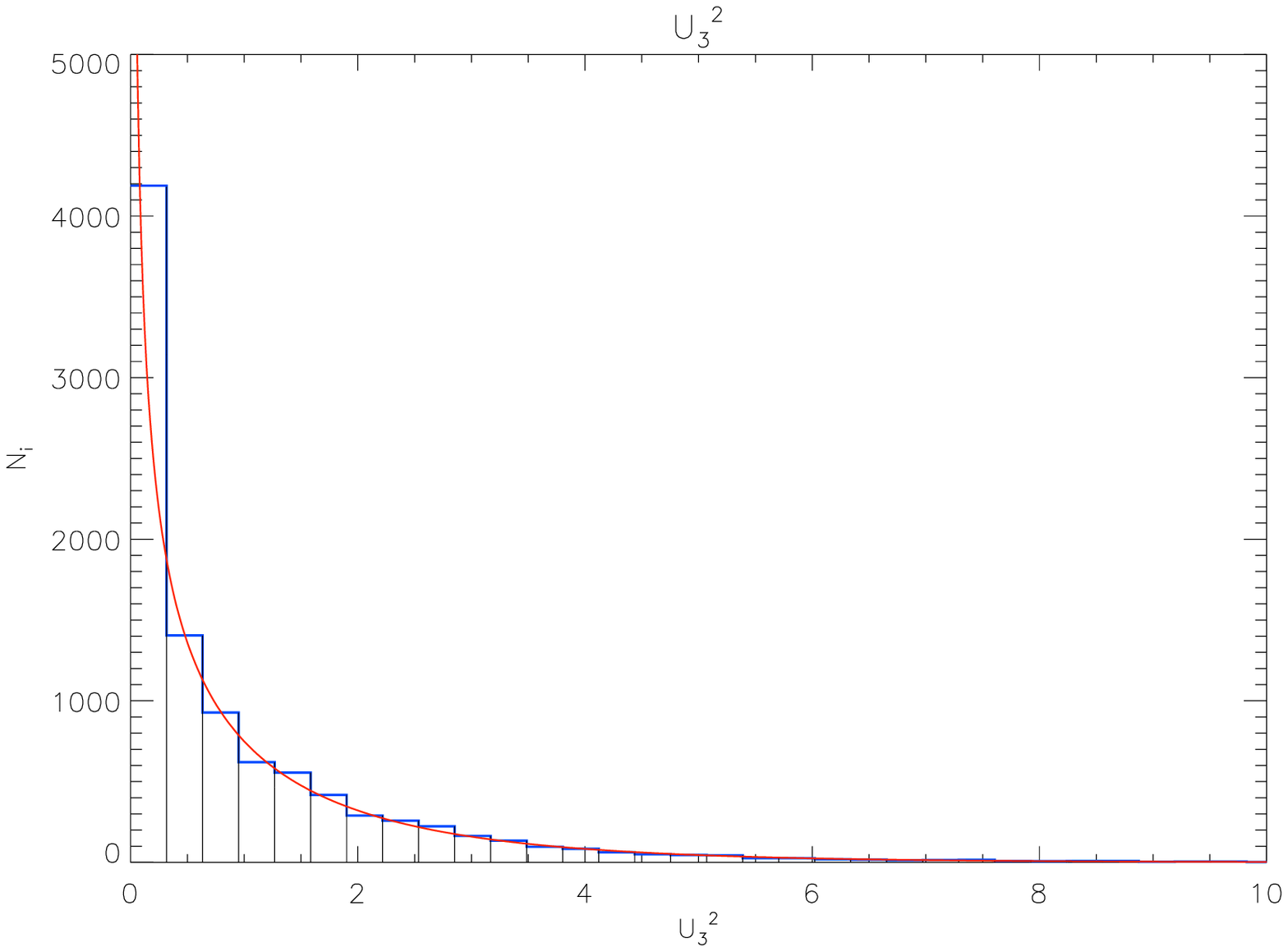}
  \includegraphics[angle=0,height=4.0cm,width=4.0cm]{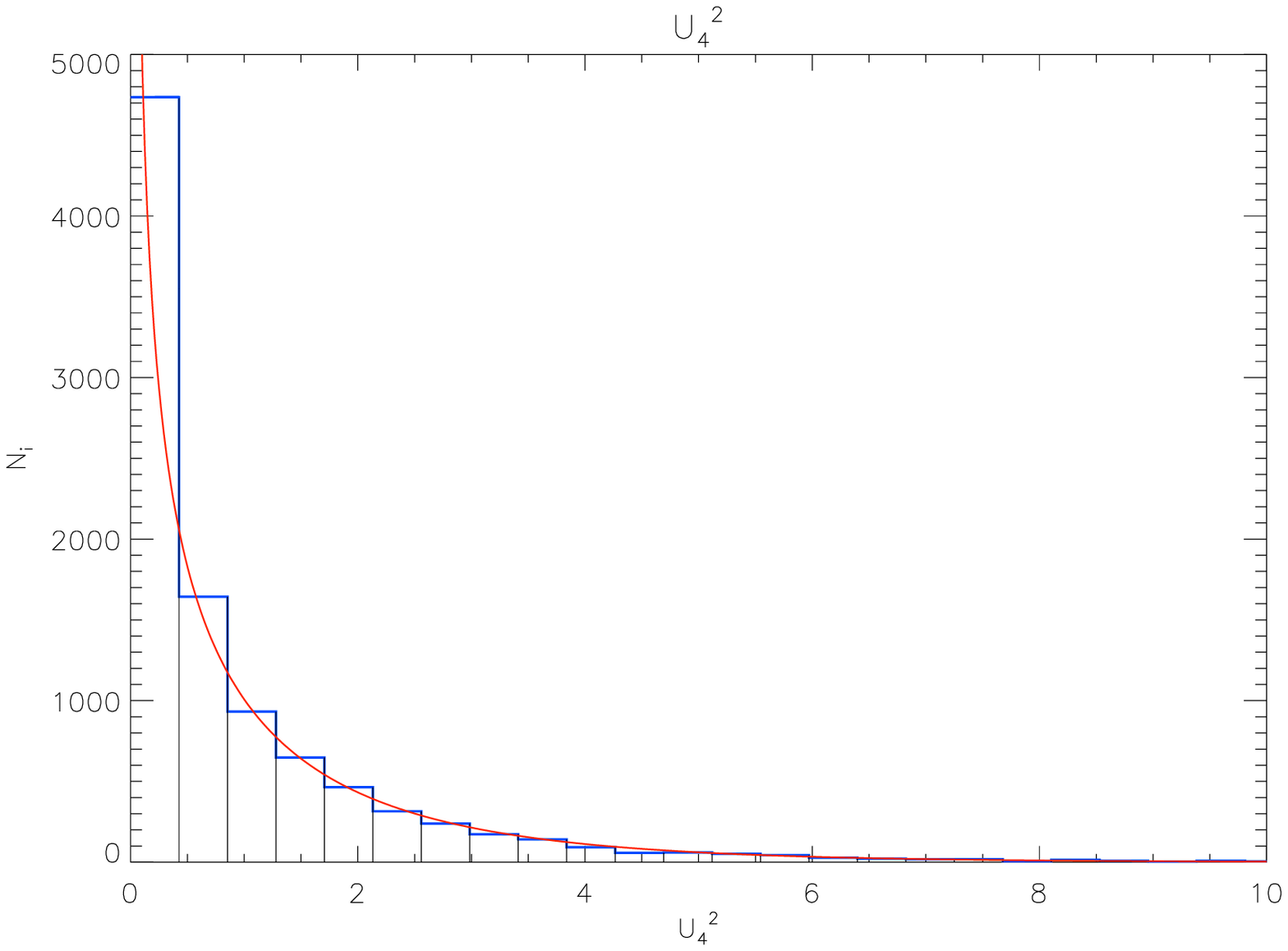}
  \includegraphics[angle=0,height=4.0cm,width=4.0cm]{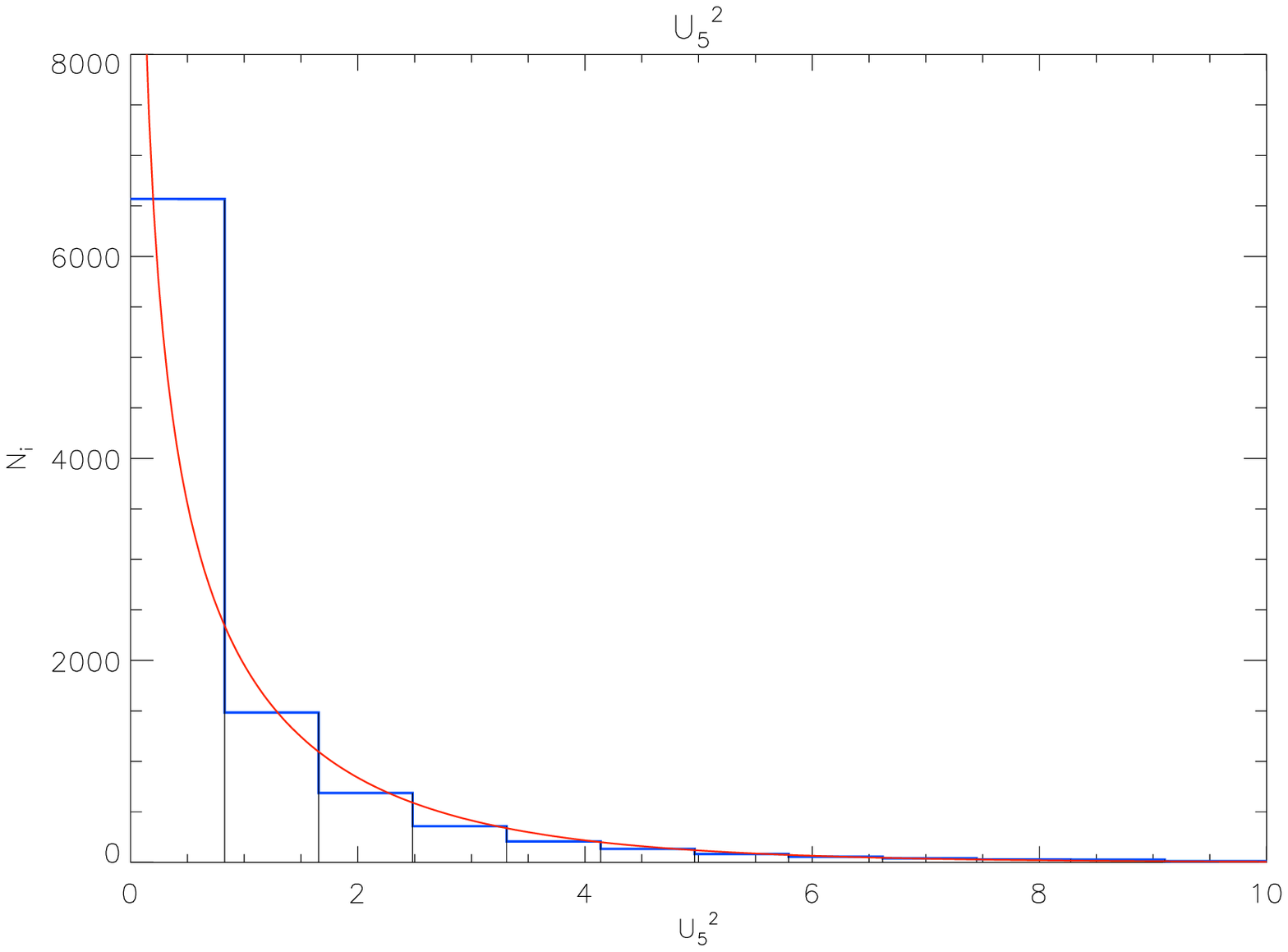}
\caption{{\it \textbf{From left to right,}} and {\it \textbf{from top
to bottom,}} distribution of the $U_i^2$ statistics, from a set of
$10^4$ Gaussian Mirage simulations analysed in the same region than
the data (figure \ref{skycoverage}).  The signal-to-noise cut which
has been used is $(s/n)_c = 0.30$.  Solid lines are the theoretical
distribution ($\chi_1^2$) normalized to the number of simulations and
the size of the binned cell. \label{histo_mirage}}
\end{center}
\end{figure*}
\begin{figure*}[tb]
 \begin{center}
   \includegraphics[angle=0,height=4.0cm,width=4.0cm]{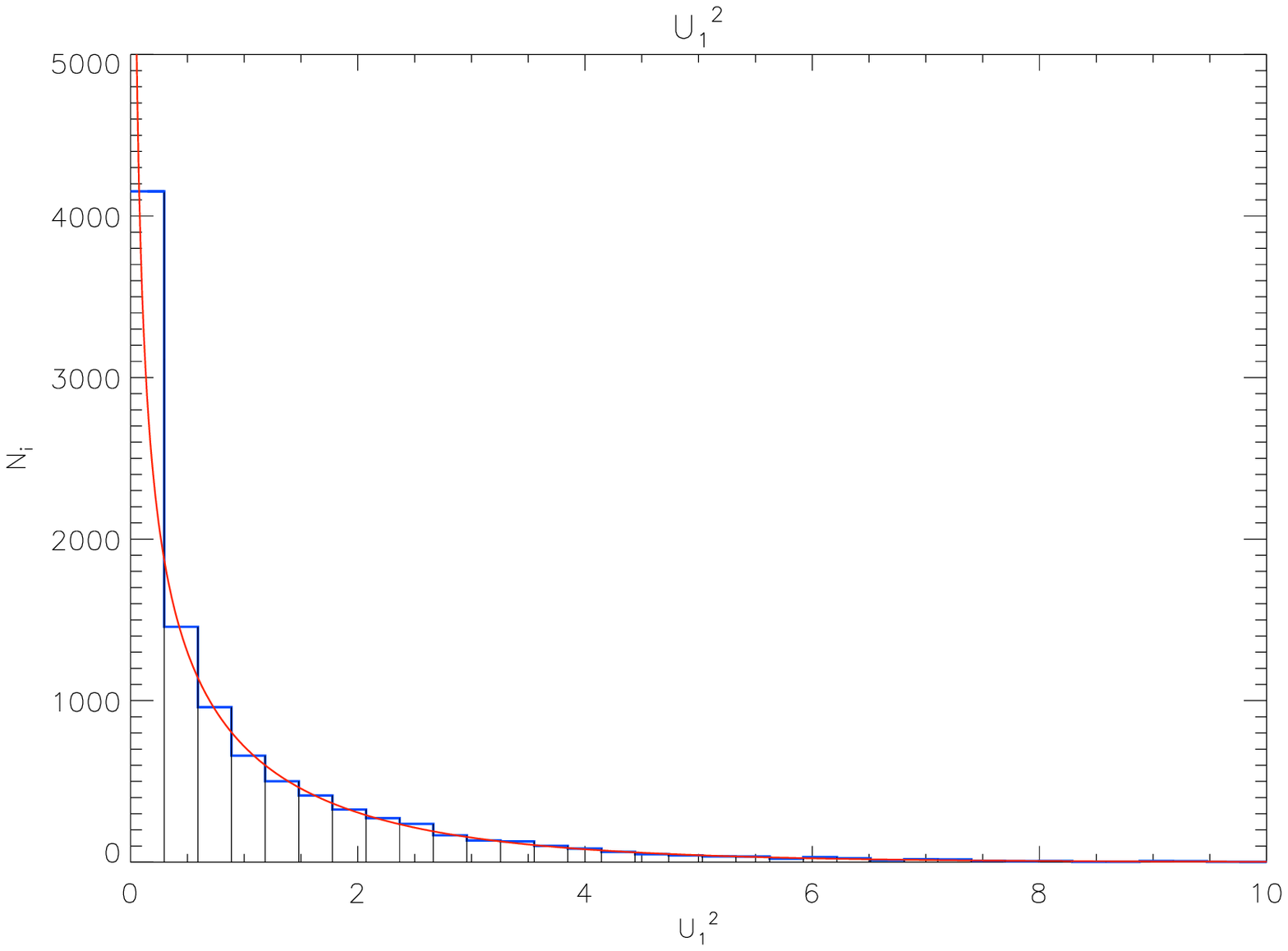}
   \includegraphics[angle=0,height=4.0cm,width=4.0cm]{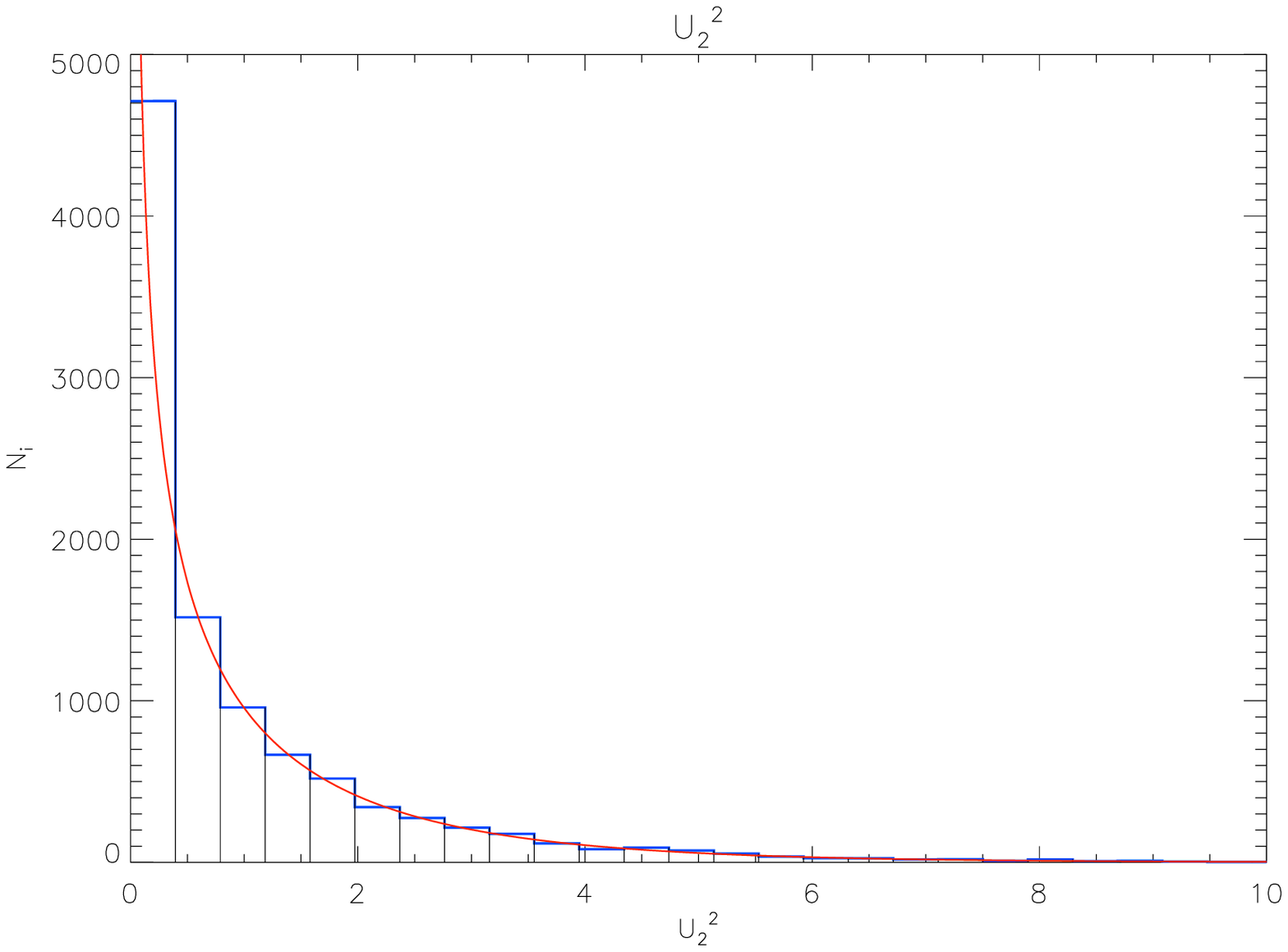}
~~
   \includegraphics[angle=0,height=4.0cm,width=4.0cm]{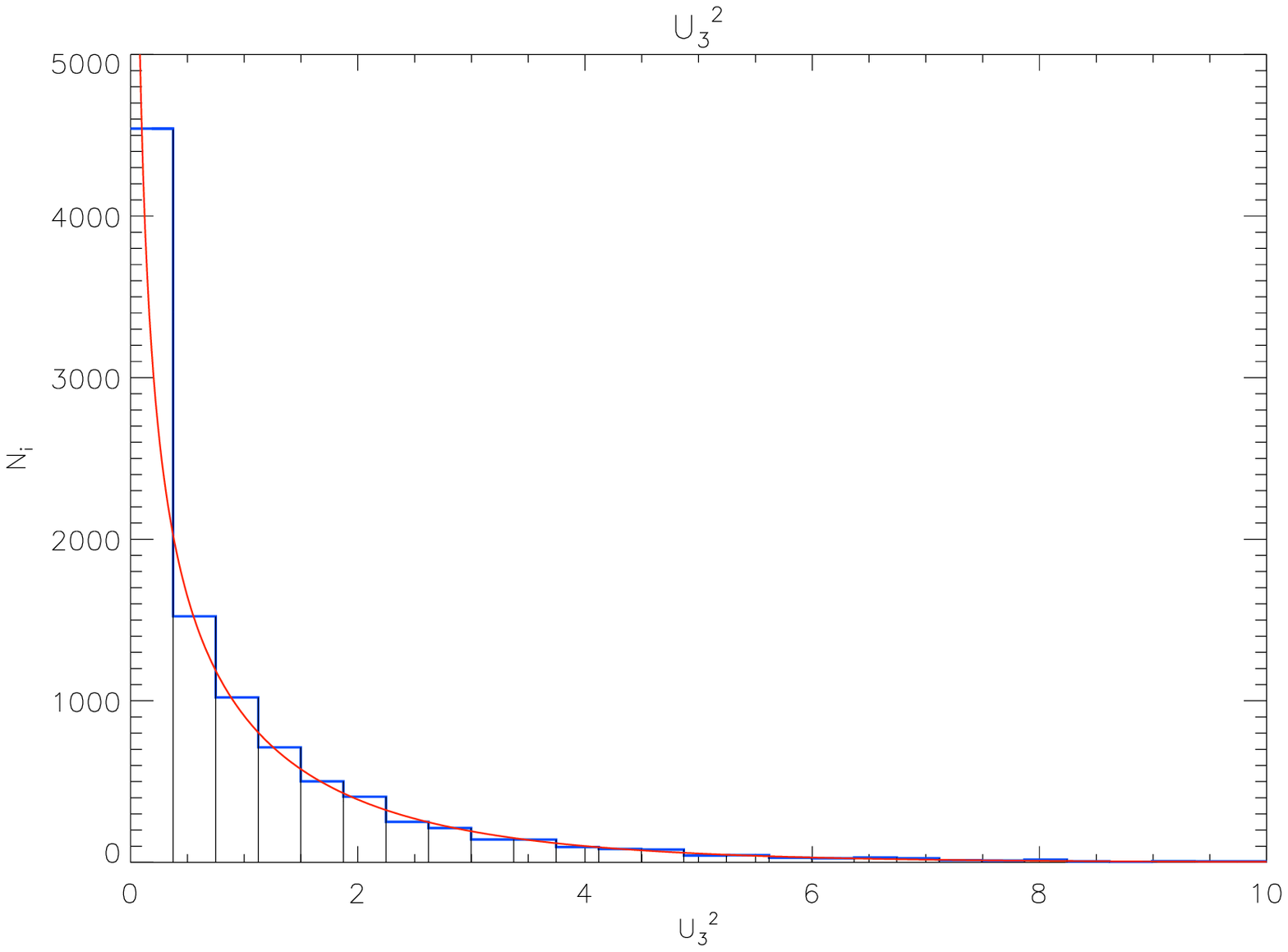}
   \includegraphics[angle=0,height=4.0cm,width=4.0cm]{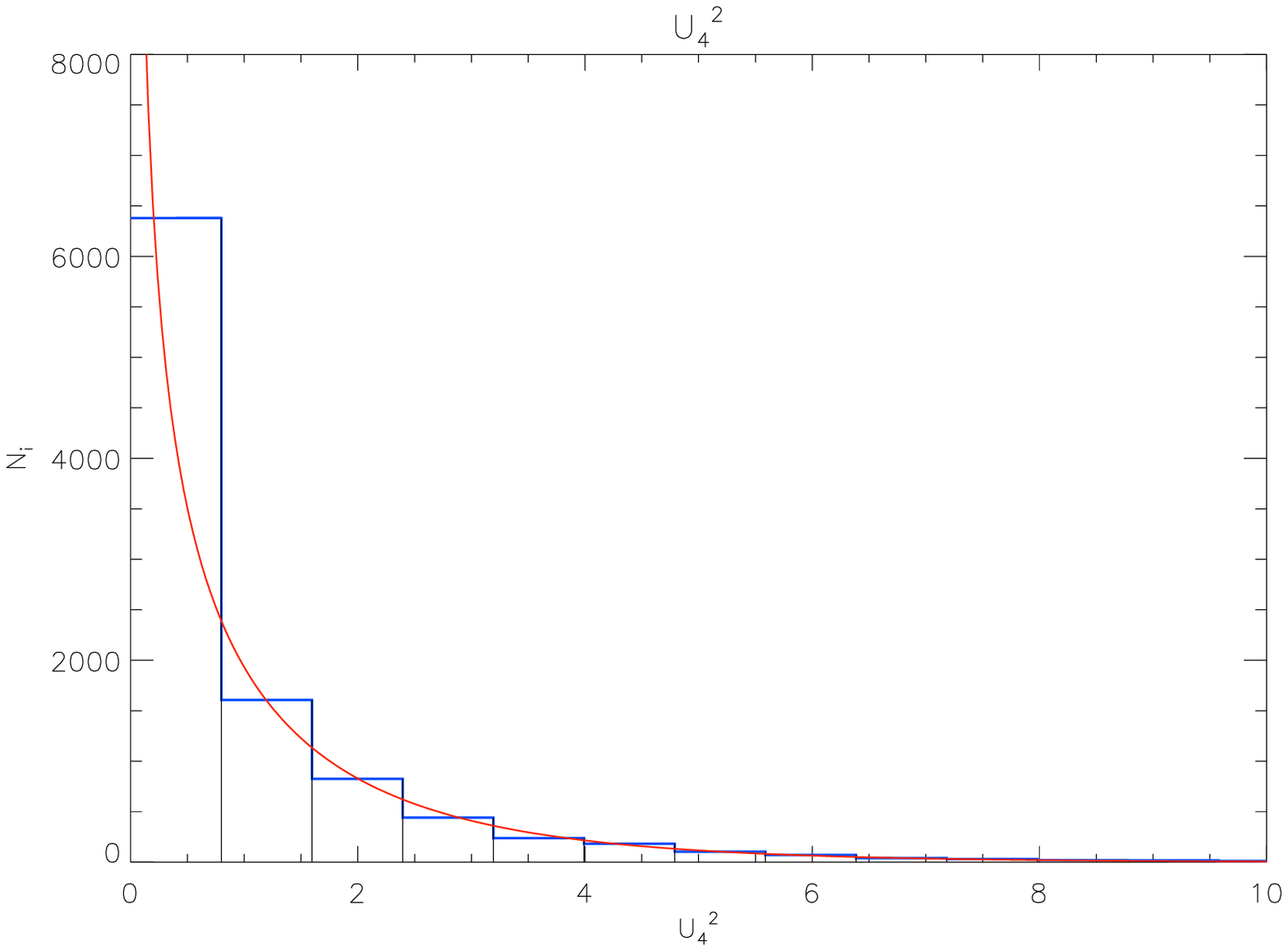}
~~
   \includegraphics[angle=0,height=4.0cm,width=4.0cm]{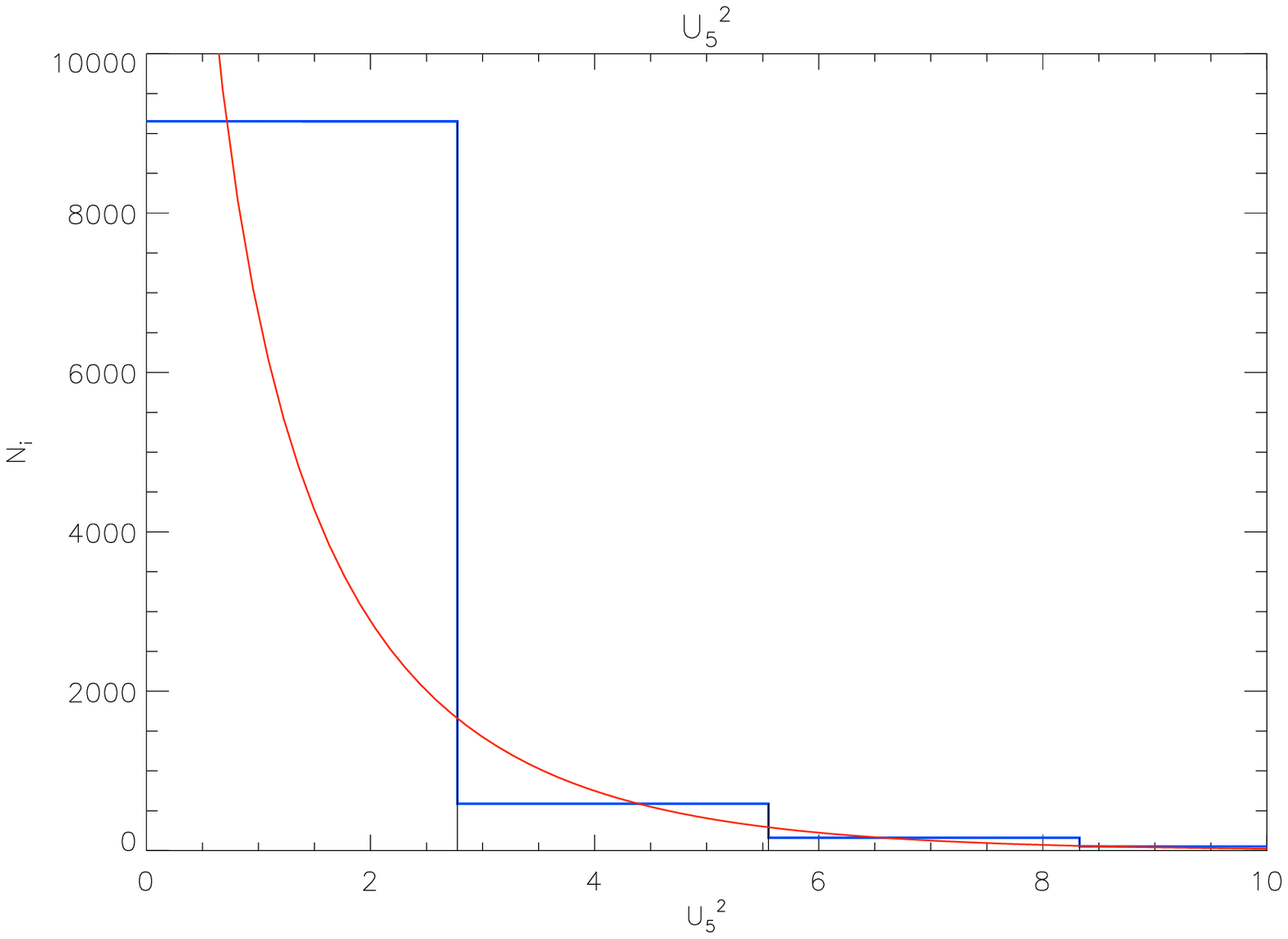}
\caption{{\it \textbf{From left to right,}} and {\it \textbf{from top
to bottom,}} distribution of the $U_i^2$ statistics, from a set of
$10^4$ Gaussian coaddition simulations analysed in the same region
than the data (figure \ref{skycoverage}).  The signal-to-noise cut
which has been used is $(s/n)_c = 0.27$.  Solid lines are the
theoretical distribution ($\chi_1^2$) normalized to the number of
simulations and the size of the binned cell. \label{histo_coadd}}
\end{center}
\end{figure*}
\begin{figure*}[tb]
 \begin{center}
   \includegraphics[angle=0,height=4.0cm,width=4.0cm]{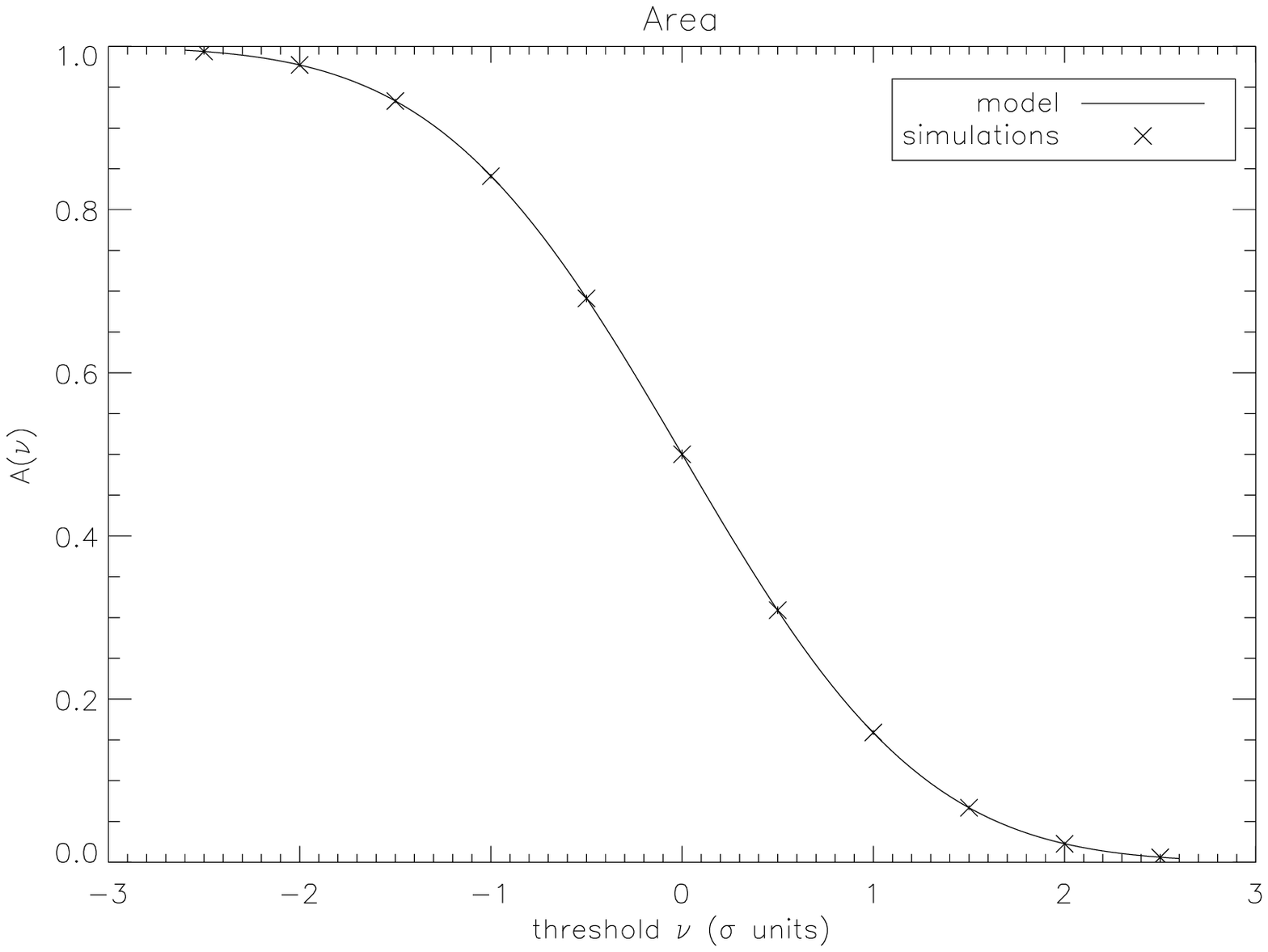}
   \includegraphics[angle=0,height=4.0cm,width=4.0cm]{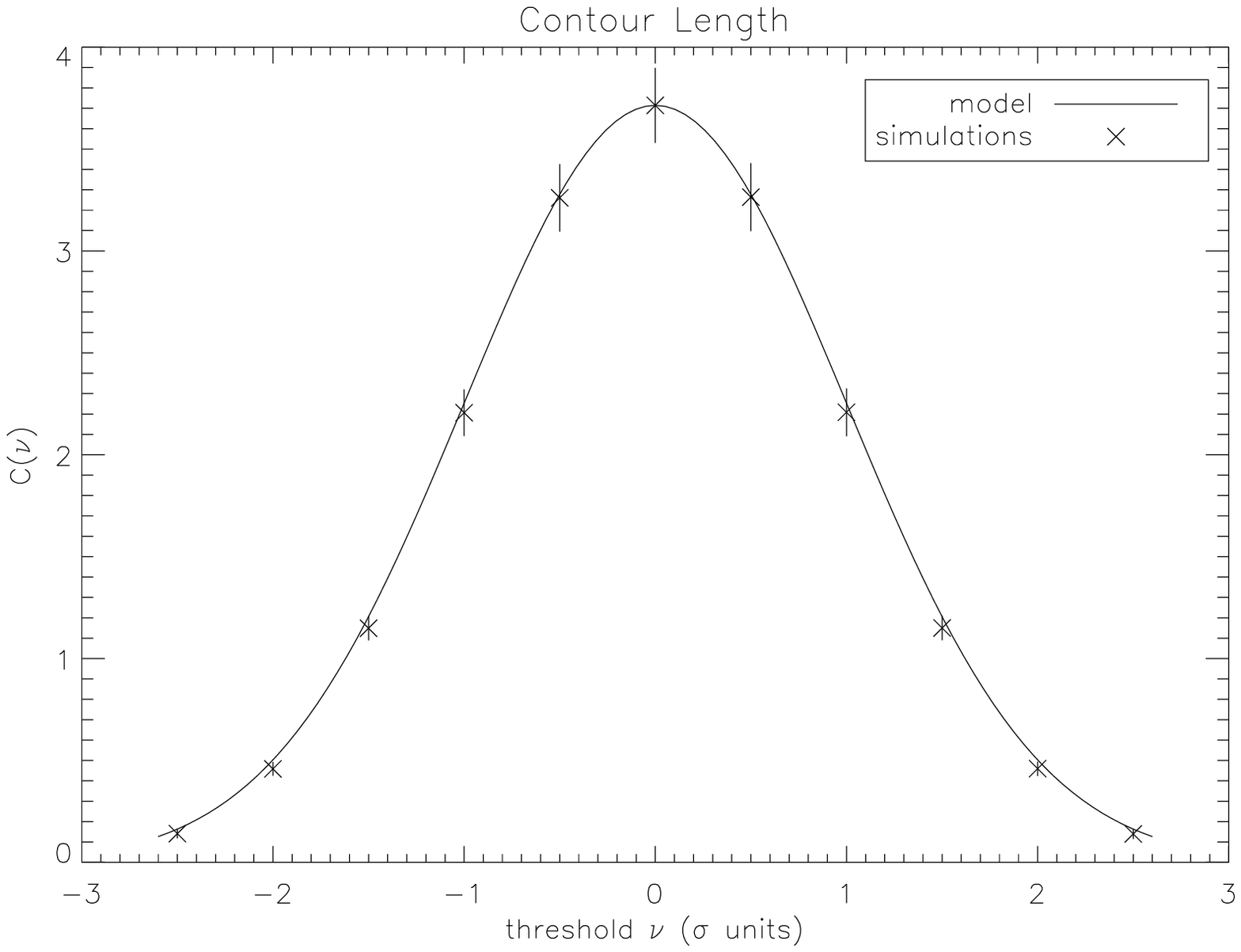}
   \includegraphics[angle=0,height=4.0cm,width=4.0cm]{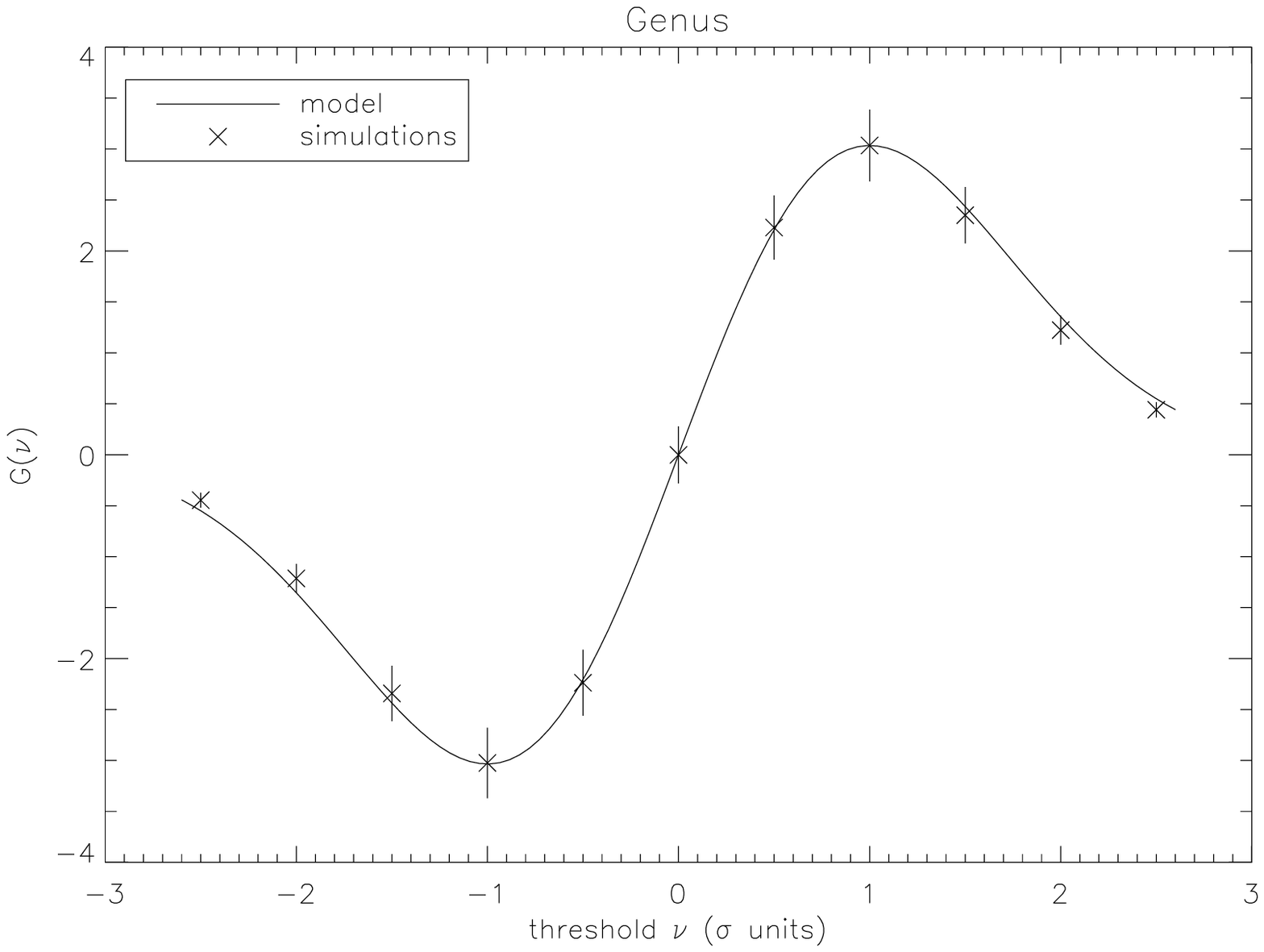}
\caption{{\it \textbf{From left to right,}} mean values of the three
Minkowski functionals and their corresponding error-bars for a set of
1000 (noiseless) CMB Gaussian simulations.  These simulations have
been generated using Archeops best fit power spectrum and have not
been masked.  Notice the good agreement between the theoretical
predictions and the results obtained from simulations.  }
\label{minkowski_cmb}
\end{center}
\end{figure*}
To develop the R\&BT non-Gaussianity test, it is necessary to
calculate the signal (S) and the noise (N) correlation matrices among
the selected pixels. We computed these matrices averaging on
simulations by means of equation \ref{corrmatrix}.  For this purpose
Monte Carlo Gaussian simulations of Archeops CMB signal and
instrumental noise were produced.  The number of performed simulations
for the map generated with the Mirage map-making procedure were $2.86
\times 10^5$ for the signal and $2.75 \times 10^5$ for the noise,
whereas for the coaddition procedure they were $5 \times 10^5$ and $5
\times 10^5$ for the signal and noise respectively. 90 dual-core 3.2
GHz processors from the IFCA computing facilities were used.  Each
Mirage simulation took 180 s of real CPU time and 1.0 GB of RAM
memory, whereas these values were 70 s and 0.04 GB respectively for
each coaddition simulation.

The high number of simulations and the corresponding computational
requirements were needed to achieve convergence in the construction of
the correlation matrices.  The main reason for the low convergence
relies on the specific properties of our correlation
matrices. Archeops noise is correlated at large scales, which means
that the $N$ matrix is neither diagonal nor sparse. The Archeops
signal correlation matrix contains correlations at large scales for
which the convergence is much slower than for the small scales due to
the cosmic variance.  In both cases many simulations ($\sim 10^5$)
were required in order to compute these matrices.

One way to quantify the degree of convergence of these matrices is by
analysing Gaussian simulations. The $U_i^2$ statistics for a set of
Gaussian simulations should have a $\chi_1^2$ pdf.  This can be
tested, for example, by calculating the mean and the variance of the
$U_i^2$ statistics for $10^4$ Gaussian signal plus noise simulations.
For the Gaussian case, the mean should be equal to 1 and the
dispersion equal to $\sqrt{2}$ (this is the null hypothesis, $H_0$).

Following \citet{aliaga_vsa,rubino}, the $U_i^2$ are computed for a
subset of signal-to-noise eigenmodes which are those associated with
eigenvalues of the signal-to-noise matrix $A$ satisfying $(D_A)_i \ge
(s/n)_c^2$, where $(s/n)_c$ is a given signal-to-noise ratio cut.  In
figure \ref{stncuts_vs_n} the number of eigenmodes $\{y_i\}$'s, which
obey $(D_A)_i \ge (s/n)_c^2$, in terms of $s/n$ is plotted.

In figure \ref{mu_stat_simu_mirage} we show the mean and dispersion of
the five first $U_i^2$ statistics for different signal-to-noise cuts
corresponding to all possible eigenvalues of the $A$ matrix.  The
$U_i^2$ values come from a set of $10^4$ Gaussian Archeops signal plus
noise Mirage simulations.  It can be seen that for $(s/n)_c \lsim 2$
mean values are close to $1$ and the dispersion close to $\sqrt{2}$
(except for the $U_5^2$ statistic whose dispersion is always larger
than 2).  As shown by e.g. \citet{aliaga_vsa}, the expected value of
$U_i^2$ is equal to 1 independently of the number of $\{y_i \}$
used. This explains why we have got the mean of $U_i^2$ very close to
$1$ for every signal-to-noise cut.  The dispersion is equal to $\sqrt
2$ asymptotically, when the number of $\{y_i \}$ used is high. In our
case, this happens for low signal-to-noise cuts, when enough 
$\{y_i \}$'s are used to compute the statistics.  In figure
\ref{mu_stat_simu_coadd} the same quantities have been plotted for the
$10^4$ Gaussian Archeops signal plus noise coaddition
simulations. Similar conclusions can be derived in this case. Notice,
however, that the results are closer to theoretical values when the
analysis is performed using the Mirage maps. In this case the
correlation matrices have converged with less simulations than in the
coaddition case. This is one of the advantages of using Mirage
simulations over the coaddition ones, although the production of a
Mirage map requires more CPU time and RAM memory than a coaddition
map.

Since the computation of high order $U_i^2$ statistics involve high
powers of the eigenmodes, the convergence of their dispersion to the
theoretical values at a given $(s/n)_c$ is slower than for the low
order ones (as can be seen in the right panels of figures
\ref{mu_stat_simu_mirage} and \ref{mu_stat_simu_coadd}).

A more exhaustive check for the convergence of the $U_i^2$ statistics
is done by comparing their theoretical pdf with the histograms
obtained from the simulated data.  Given a signal-to-noise ratio cut
$(s/n)_c$ for the calculation of the $U_i^2$ statistics, it is
possible to make a histogram with the corresponding values of the
$U_i^2$ statistics from the same sets of $10^4$ simulations.  Figure
\ref{histo_mirage} compares the histograms for the first five
statistics calculated using all the eigenmodes ($s/n \ge 0.30$) for
the Mirage simulations with the theoretical expectation of a
$\chi_1^2$ distribution. In table \ref{table_stat_mirage} the mean and
the dispersion of these histograms are presented. In figure
\ref{histo_coadd} the same comparison is shown for the coaddition
simulations considering also all the eigenmodes ($s/n \ge 0.27$). The
corresponding mean and dispersion of these histograms are given in
table \ref{table_stat_coadd}.

In summary, the four statistics $U_1^2$, $U_2^2$, $U_3^2$, and $U_4^2$
have a pdf's compatible with the theoretical one whereas
$U_5^2$ starts to deviate from it. The discrepancy, already present in
the dispersion, increases for higher orders. The reason is that high
order moments enlarge possible errors present in the computed
correlation matrices and are propagated in the diagonalization
processes.  In any case, the $U_5^2$ statistic can still be used for
the Gaussian analysis if the distribution obtained from the
simulations, instead of the theoretical one, is used as
reference.  Although this is not as optimal as using the theoretical
$\chi_1^2$ distribution, it is however a good compromise taking into
account the huge computational resources needed to produce a very
large number of simulations.

For the Minkowski functionals analysis the expected values
given by equation \ref{area} cannot be applied to our problem because
of the contour restrictions of the mask and the presence of
anisotropic noise. Nevertheless in order to test our Minkowski
functional codes we performed an analysis on (noiseless) CMB Gaussian
simulations over all sky and 1.8 degrees resolution generated using the
best fit Archeops power spectrum. Analysing them for thresholds from
$ -2.5\sigma$ to $ 2.5\sigma$ (where $ \sigma$ is the standard deviation
of the corresponding simulation), we obtained that the results from
simulations are compatible with the theoretical predictions (see
fig. \ref{minkowski_cmb}).
\begin{table}[t]
  \center
  \caption{Mean and dispersion of $U_i^2$ statistics from $10^4$ Mirage
    simulations for a signal-to-noise ratio cut of 0.30.
    \label{table_stat_mirage}}
  \begin{tabular}{c|ccccc|c}
    \hline \hline
    ... & $U_1^2$ & $U_2^2$ & $U_3^2$ & $U_4^2$ & $U_5^2$ &  $\chi_1^2$ \\
    \hline
    $\mu$ & 1.02 & 1.04 & 1.01 & 1.01 & 1.02 &  1.00 \\
    $\sigma$ & 1.45 & 1.47 & 1.43 & 1.55 & 1.96 & 1.41 \\
    \hline
    \hline
  \end{tabular}
  \caption{Mean and dispersion of $U_i^2$ statistics from $10^4$ coaddition 
    simulations for a signal-to-noise ratio cut of 0.27.
    \label{table_stat_coadd}}
  \begin{tabular}{c|ccccc|c}
    \hline \hline
    ... & $U_1^2$ & $U_2^2$ & $U_3^2$ & $U_4^2$ & $U_5^2$ &  $\chi_1^2$ \\
    \hline
    $\mu$ & 0.99 & 1.02 & 1.02 &  1.02 & 1.00 &  1.00 \\
    $\sigma$ & 1.40 & 1.47 & 1.48 & 1.62 & 2.27 &  1.41 \\
    \hline
    \hline
  \end{tabular}
\end{table}
\section{Gaussianity test on Archeops data}
We have applied the R\&BT to the Archeops $143K03$ bolometer map. The
signal-to-noise eigenmodes have been computed with the correlation
matrices described in section \ref{section4}, for each map-making
case. We have checked in that section that these signal and noise
matrices provide $U_i^2$ statistics compatible with Gaussianity for
Gaussian simulations.

\begin{figure}[t]
\center
\epsfig{file=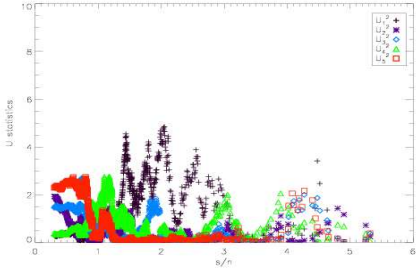,height=3.8cm,width=6.0cm}
\caption{$U_i^2$ statistics of Mirage Archeops Data for different signal-to-noise cuts.}
\label{fig:archeops_1995_ui}
\end{figure}

\begin{figure}[t]
\center
\epsfig{file=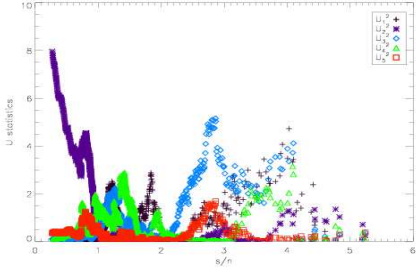,height=3.8cm,width=6.0cm}\
\caption{$U_i^2$ statistics of coaddition Archeops Data for different signal-to-noise cuts.}
\label{fig:coaddarcheops_1995_ui}
\end{figure}

We have applied this test to the Archeops data for the Mirage and
coaddition map-making. The $U_i^2$ statistics, computed for the 1995
pixels of the previously described Archeops data, are displayed
in figures \ref{fig:archeops_1995_ui} and
\ref{fig:coaddarcheops_1995_ui}. The $U^2_i$ statistics are plotted,
from i = 1 to 5, versus the signal-to-noise eigenmode cut.

For the Mirage map-making, results are displayed on figure
\ref{fig:archeops_1995_ui}. We can see that all the $U_i^2$ statistics
are below 5 for all the signal-to-noise cuts. This means that the data
is compatible with Gaussianity.

For coaddition map-making, we can see from figure
\ref{fig:coaddarcheops_1995_ui} that whatever the signal-to-noise
eigenmode cut is, $U_i^2$ statistics for the 143K03 bolometer data are
below 5, except for $U_2^2$ for signal-to-noise cuts below 0.5.  It
reaches the maximum value of 7.97 at the minimum signal-to-noise cut
of 0.27.  The upper tail probability
\footnote {$\int_a^{\infty}f(y)dy$}
for $U_2^2=7.97$ from the
$\chi_1^2$ distribution (equation \ref{pdfchi}) is 0.5\%. Comparing
with the set of coaddition Gausian simulations we found that this
upper tail probability is 0.6\%, (see table \ref{tabcoadd}), in good
agreement with the theoretical expectation.  Nevertheless, as we
have computed $U_i^2$ statistics for all possible signal-to-noise
cuts, it is important to estimate the significance of finding any
simulation with $U_2^2 \geq 7.97$ in at least one of them. This is the
so called ``p-value'' of $U_2^2$. The ``p-value'' is defined as the
probability that the relevant statistic takes a value at least as
extreme as the one observed by the data when the null hypothesis is
true. We have found for $U_2^2$ that the ``p-value'' is 15.0\%.

Then we can conclude that even if we have a relatively strong $U_2^2$
at the lowest signal-to-noise ratio, it is not unlikely to have
such a high value by chance. Therefore, even considering the results
from the coaddition map-making, Archeops data is still compatible with
our Gaussian simulations.

Although the high value found for $U_2^2$ for the coaddition map is
not significant enough to be incompatible with Gaussianity, it is
clear that there is a steady increase of $U_2^2$ when $s/n$
decreases. This suggests the presence of systematics in the coaddition
maps that can depend on the resolution.  Moreover, the fact that it
only appears in coaddition data suggests the possibility that it is a
map-making issue.  This also implies that systematics are better
controlled in the Mirage than in the coaddition map-making.
Therefore hereafter we focus only on the Mirage map making
data.
\begin{table}[t]
  \center
  \caption{
    $U_i^2$ from Archeops Mirage (coaddition) map for $(s/n)_c=0.30$
    ($(s/n)_c=0.27$) and the probability that one $s+n$ Gaussian Mirage
    (coaddition) simulation has a $U_i^2$ statistic larger than
    those of data. More precisely, the probability for $U_2^2$ in 
    the coaddition case is 0.6\%.
  }
  \label{tabcoadd}
  \begin{tabular}{c|ccccc}
    \hline \hline
    ... & $U_1^2$ & $U_2^2$ & $U_3^2$ & $U_4^2$ & $U_5^2$ \\
    \hline
    \textbf{Mirage} & 0.28 & 1.92 & 1.45 & 0.38 & 2.29 \\
    \textbf{Prob.} & 0.60 & 0.17 & 0.23 & 0.54 & 0.12 \\
    \hline
    \textbf{Coaddition} & 0.11 & 7.97 & 0.10 & 0.04 & 0.34 \\
    \textbf{Prob.} & 0.73 & 0.01 & 0.75 & 0.83 & 0.52 \\
    \hline
    \hline
  \end{tabular}
\end{table}

\begin{figure}[t]
\center
\epsfig{file=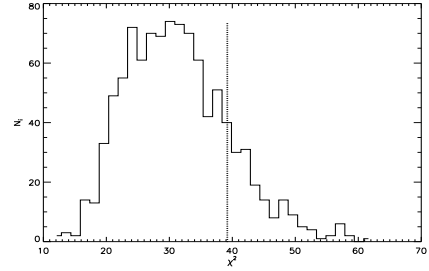,height=3.8cm,width=6.0cm}
\caption{Distribution of the $\chi^2$ values from the Minkonwski
    Gaussianity test for Archeops Mirage map. Vertical line shows the
    data results. Their cumulative probability is 83.9\%.}
\label{mkinkowskchi}
\end{figure}

We performed a $ \chi^2$ test with the three Minkowski functionals
using 11 thresholds from $ -2.5\sigma$ to $ 2.5\sigma$. We analysed
the Mirage data and a set of 1000 CMB Gaussian simulations with noise
of the Mirage type. The corresponding histogram of the $ \chi^2$
values of these simulations and of the data are presented in figure
\ref{mkinkowskchi}. As it can be seen, the data are compatible with
the Gaussian simulations.
\subsection{Systematic and foreground contamination}
\begin{figure}[bt]
\center
\epsfig{file=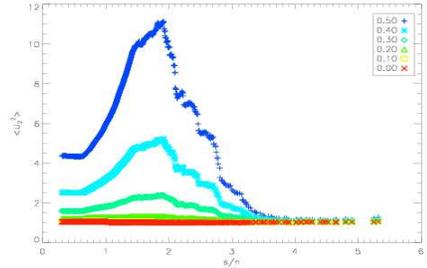,height=3.8cm,width=6.0cm}
\caption{Mean of $10^4$ $U_2^2$ statistics, from $10^4$ signal plus
noise Mirage simulations plus a factor $\alpha_d$ times the
contamination template.  $0.0 \le \alpha_d \le 0.5$.}
\label{fig:dust_1995_u2_mirage}
\end{figure}
\begin{figure*}[bt]
\begin{center}
\epsfig{file=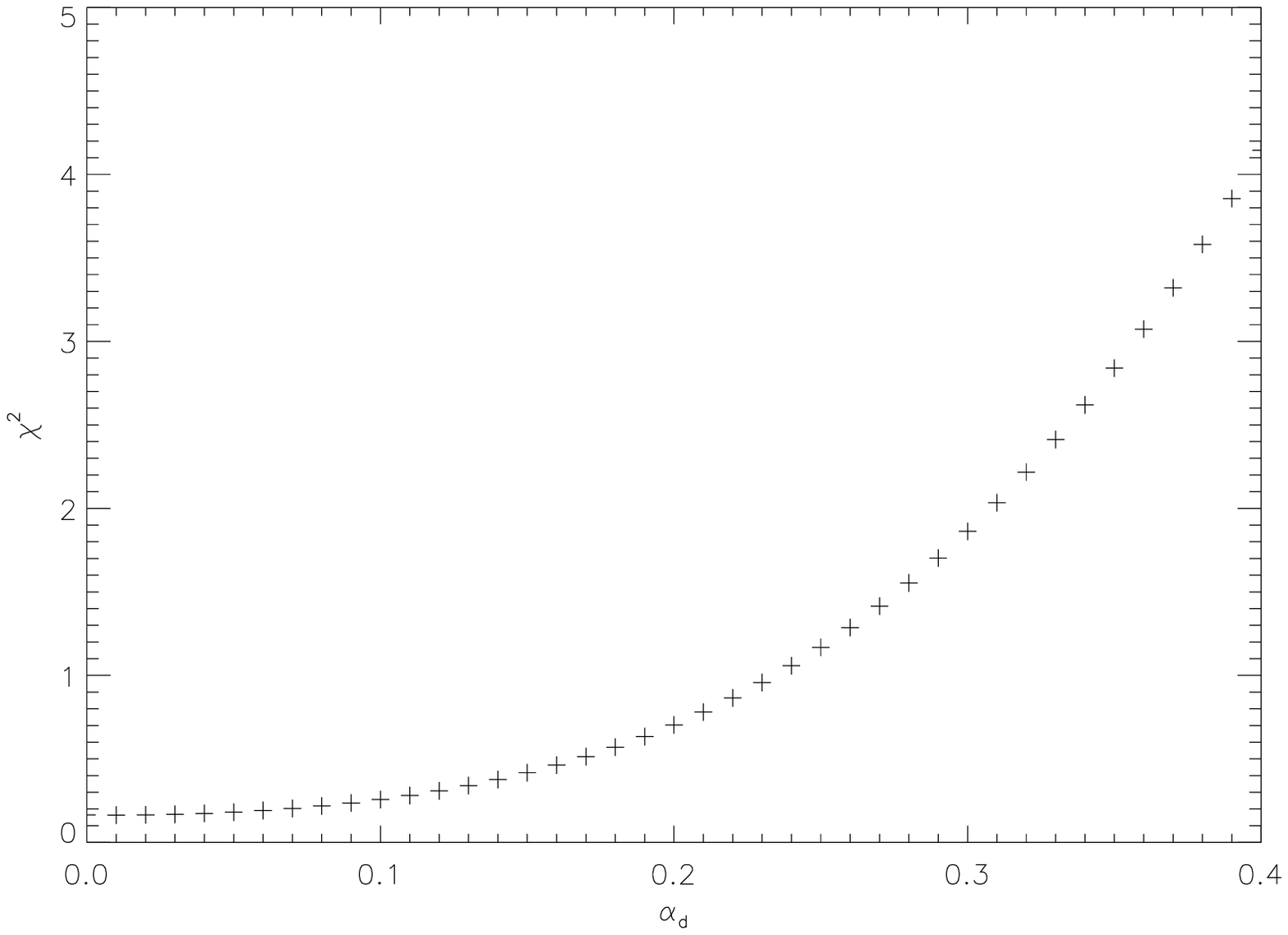,height=3.8cm,width=6.0cm}
\epsfig{file=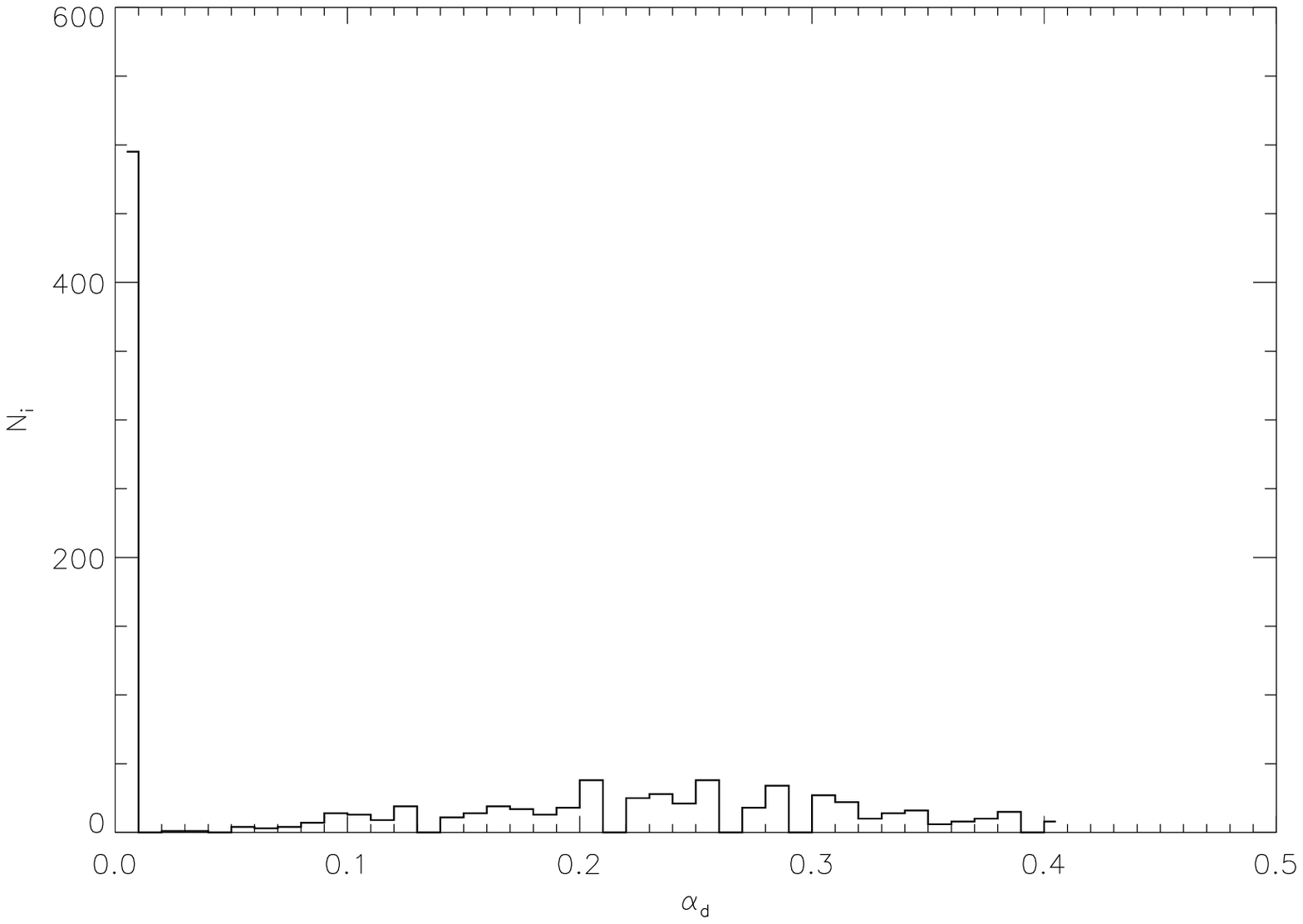,height=3.8cm,width=6.0cm}
\caption{{\it \textbf{From left to right,}} $\chi^2$ value of Archeops
data for different $ \alpha_d$ and the histogram of best fit $
\alpha_d$ for a set of 1000 Gaussian simulations without dust. These
results have been obtained with the $ U_2$ statistic.}
\label{chidust}
\end{center}
\end{figure*}
The R\&BT can also provide a powerful tool for estimating the level of
this contribution. The test consists in adding different
percentages of a template map to the Archeops 143K03 bolometer map,
for the Mirage and coaddition simulations cases, to compare the
resulting $U^2_i$ statistics to the ones obtained with the Archeops
data at 143 GHz.

This template map is computed from the coadded Archeops 353 GHz map
\citep[see][]{ponthieu05}. This map contains thermal dust emission,
atmospheric residuals as the dominant components and also
instrumental noise and CMB residuals.  Thus, extrapolated to 143 GHz
it will provide a good template of what could be a dust plus
atmospheric contamination at this frequency.

Thermal dust is assumed to have a grey-body emission: $\nu^2B(\nu)$
which can be approximated in the Rayleigh-Jeans domain to 
$T_{\rm RJ}\propto \nu^2$ \citep[see][]{ponthieu05}.  Atmospheric residuals
emission law has been estimated empirically by the Archeops
collaboration \citep[see][]{archeops_tecnical} and is also 
proportional to $\nu^2$ in the Rayleigh-Jeans domain.  Dust and
atmospheric residuals being the two main components, Archeops 353 GHz
map has been extrapolated to 143 GHz by assuming that emission
power law. Due to the extrapolation the CMB contribution on the
353 GHz template map is negligible with respect to the CMB at 143 GHz.

$U_2^2$ statistic is the most sensitive to this effect as can be
seen in figure \ref{fig:dust_1995_u2_mirage} for the Mirage case where
this statistic presents a prominent peak at signal-to-noise ratio cuts
around 1.88.

In order to determine the level of contamination we performed
a $ \chi^2$ test with the $ U_2$ statistic computed at $
(s/n)_c = 1.88$. It is optimal to perform a $ \chi^2$ test with
$ U_2$ because $ U_2$ is normally distributed for the null
hypothesis. Thus we can define
\begin{equation}
\chi^2(\alpha_d) = \frac{1}{\sigma^2_{\alpha_d}(U_2)}
(U_2-\langle U_2 \rangle_{\alpha_d})^2
\end{equation}
where $ \langle U_2 \rangle_{\alpha_d}$ and $ \sigma_{\alpha_d}(U_2)$
are the mean and the dispersion of $ U_2$ for CMB Gaussian simulations
with noise plus a factor $ \alpha_d$ times the contamination
template. In the left panel of figure \ref{chidust} we present the $
\chi^2$ of Archeops Mirage data for different $ {\alpha_d}$. We can
see that the minimum $ \chi^2$ (best fit) occurs for $
{\alpha_d}$=0.0.  Analysing Gaussian simulations without dust we find
that most of them reach the best fit for low values of $ {\alpha_d}$
(right panel of figure \ref{chidust}).  Specifically we have that $
\alpha_d \le$ 0.27 for 90\% confidence level (CL), and $ \alpha_d \le$
0.33 for 95\% CL. By comparing the dispersion of both maps, Archeops
and 0.27 times the contamination template, we can exclude a dust plus
atmospheric contamination larger than 7.8\%.

We computed another $ \chi^2$ statistic using the Minkowski
functionals for the dust analysis. In this case
\begin{equation}
\chi^2(\alpha_d) = \sum_{i,j}(\vec v(i) - \langle \vec v(i)
\rangle_{\alpha_d})C^{-1}_{ij}(\vec v(j) - \langle \vec v(j)
\rangle_{\alpha_d})
\end{equation}\label{chialphadmink}
$i$ and $j$ cover 11 thresholds from $ -2.5\sigma$
to $ 2.5\sigma$ and the three Minkowski functionals. $ \langle
\vec v(i)\rangle_{\alpha_d}$ is the mean value of the corresponding
functional at the corresponding threshold for Gaussian CMB simulations
with noise plus $ \alpha_d$ times the dust template. $ C$ is the
covariance matrix for Gaussian CMB simulations with noise. The value
of $ \alpha_d$ that best fits Archeops data is $ \alpha_d =$
0.0. Analysing Gaussian simulations without dust we find that $
\alpha_d \le$ 0.28 for 90\% CL, and $ \alpha_d \le$ 0.35 for 95\%
CL.
\subsection{Primordial non-Gaussianity}
\begin{figure*}[bt]
\begin{center}
\epsfig{file=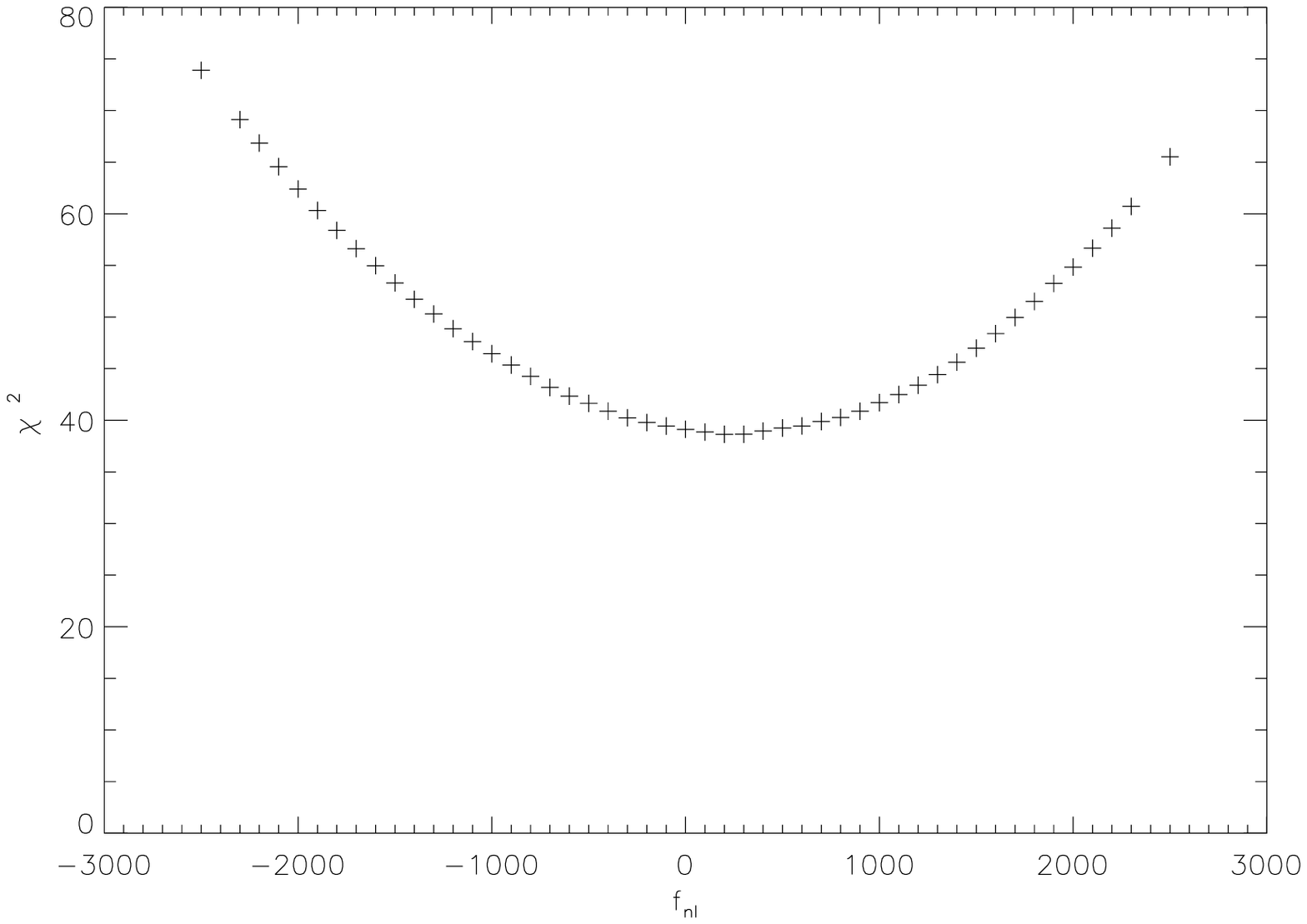,height=3.8cm,width=6.0cm}
\epsfig{file=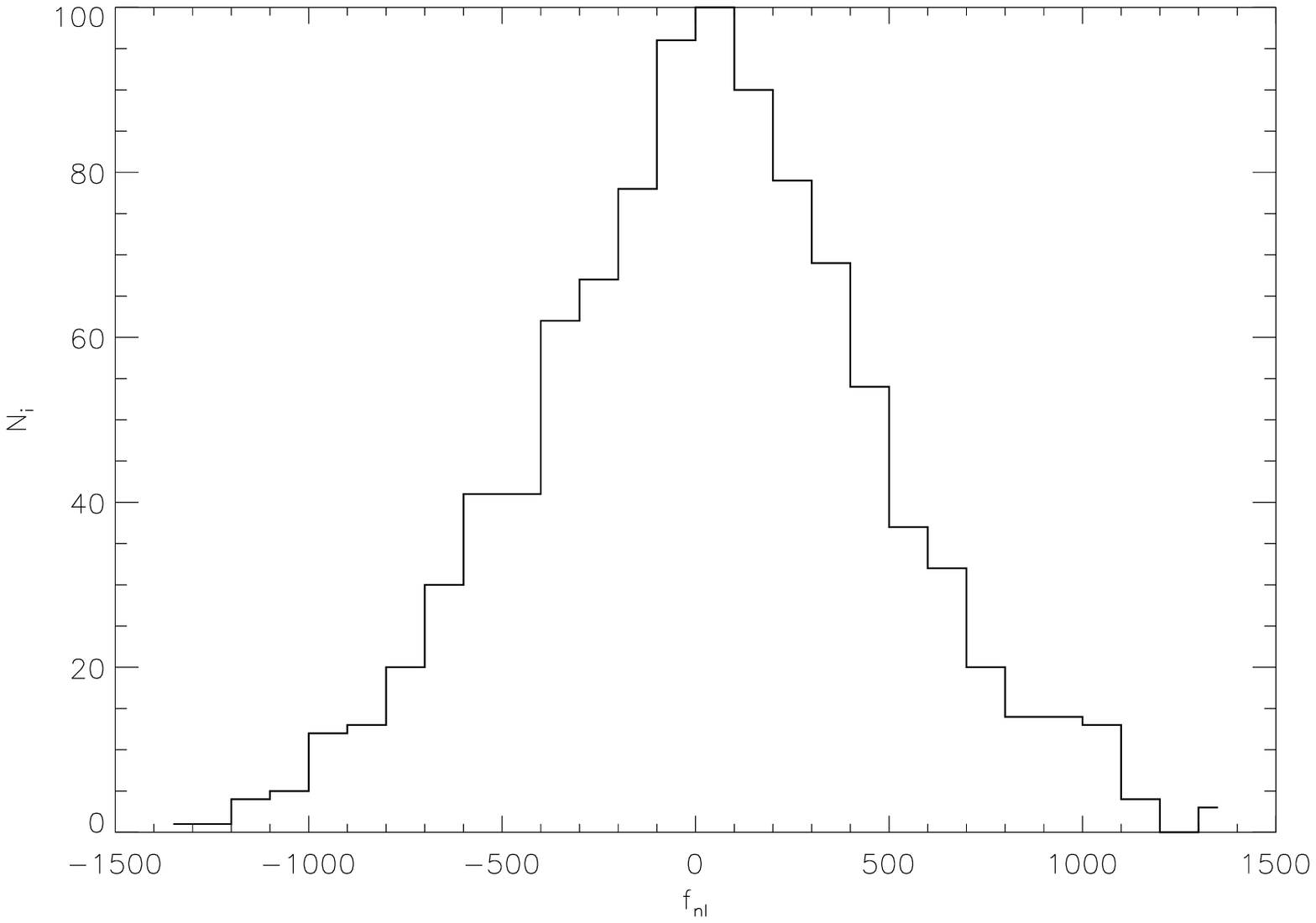,height=3.8cm,width=6.0cm}
\caption{{\it \textbf{From left to right,}} $\chi^2$ value of Archeops
data for different $ f_{nl}$ and the histogram of best fit $ f_{nl}$
for a set of 1000 CMB Gaussian simulations with noise. These results
have been obtained with the Minkowski functionals.}
\label{chifnl}
\end{center}
\end{figure*}
There are several possible inflationary scenarios in which the
primordial fluctuations are not Gaussian distributed.  The idea is to
work with a simple non-Gaussianity model and to impose some
constraints on it.  In particular, we consider the ``weak non-linear
coupling case'' \citep{komatsu2001,liguori2003,bartolo2004}
\begin{equation}
\Phi(\vec x) = \Phi_L(\vec x) + f_{nl}\{\Phi^2(\vec x)-\langle \Phi^2(\vec x) \rangle \} \ \ 
\end{equation}
where $\Phi(\vec x)$ is the primordial gravitational potential, (which 
satisfies $\langle \Phi(\vec x) \rangle = 0$), $\Phi_L(\vec x)$ is the
linear random component (Gaussian distributed), and $f_{nl}$ is the
non-linear dimensionless\footnote{We use the units system with $c = 1$.}
coupling parameter.

Scales larger than 1 degree are larger than the horizon scale at
the recombination time, when CMB was formed \citep{liddle}. In this
regime it is possible to make a good approximation linking CMB
fluctuations and gravitational fluctuations through the Sachs-Wolfe
effect \citep{sachswolfe} $\Delta T(\vec n)/T = \Phi(\vec n)/3$
(notice however that a better approximation should include the
integrated Sachs-Wolfe effect).

We analysed signal plus noise simulations with a $f_{nl}$ term in this way,
\begin{eqnarray}
\nonumber
\Delta T'_s(\vec n) & = & \Delta T_s (\vec n)+\frac{3f_{nl}}{T}\{\Delta T_s(\vec n)^2 - \langle \Delta T_s(\vec n)^2\rangle \} \\
\Delta T(\vec n) & = & \Delta T'_s(\vec n)+ \Delta T_n(\vec n) \ \ ,
\label{nogchi}
\end{eqnarray}
where $\Delta T_s$ is a Gaussian signal simulation, $\Delta T_n$ is a
Gaussian noise simulation, $T = 2.725~K$ and $\Delta T$ is the
analysed simulation.

We performed a $ \chi^2$ analysis for the primordial non-Gaussianity
similar to the dust case for both $ U_2$ and the Minkowski
functionals. The signal-to-noise eigenmodes $ y_i$ are weakly
dependent on $ f_{nl}$. It can be seen that the mean value of $ y_i^2$
for simulations with $ f_{nl}$ is
\begin{eqnarray}
\langle y_i^2 \rangle_{f_{nl}} = 1 + \frac{a_i}{1+(D_A)_i}*f_{nl}+
\frac{b_i}{1+(D_A)_i}*f_{nl}^2 \\ a_i =
\frac{1}{T}\sum_{j,k}(R_A^tL_N^{-1})_{ij}(\langle s_js_k^2\rangle + \langle
s_ks_j^2\rangle)(L_N^{-t}R_A)_{ki} \\ b_i=
\frac{1}{T^2}\sum_{j,k}(R_A^tL_N^{-1})_{ij}(\langle s_j^2s_k^2\rangle + \langle
s^2\rangle^2)(L_N^{-t}R_A)_{ki}
\end{eqnarray}
where $ b_i$ is about an order of magnitude larger than $ a_i$ for
most of the $ s/n$ eigenmodes. This implies that $ \langle y_i^2
\rangle_{f_{nl}}-1 \sim O(f_{nl}^2)$ which explains the low
sensitivity of $ U_2$ to $ f_{nl}$ variations. In particular, we have
found that it is much less sensitive than the Minkowski
functionals. If we consider for example, a value of $ f_{nl} = $ 2300,
we find a relative variation $ (\langle y_i^2\rangle_{f_{nl}} -
\langle y_i^2\rangle_{0})/ \langle y_i^2\rangle_{0} \simeq 0.05$ (and
therefore a similar ratio for $ U_2$ and $ U_2^2$) for the former and
$ (\langle F^2 \rangle_{f_{nl}} - \langle F^2\rangle_{0})/ \langle F^2
\rangle_{0} \simeq 0.50$ for the latter.

Therefore we performed a $ \chi^2$ test with the three Minkowski
functionals using different thresholds between $ -2.5\sigma$ and $
2.5\sigma$. In the left panel of figure \ref{chifnl} we present the $
\chi^2$ value of the data for different $ f_{nl}$ cases. We can see
that the minimum $ \chi^2$ value is reached for $ f_{nl}=$ 200. Taking
also into account the results obtained when analysing Gaussian
simulations (see right panel of figure \ref{chifnl}) we can put the
following constraints on $ f_{nl}$ from the Archeops data: $ f_{nl} =
200_{~-300}^{~+600}$ at 68\% CL, $ f_{nl} = 200_{~-600}^{~+900}$ at
90\% CL, and $ f_{nl} = 200_{~-800}^{~+1100}$ at 95\% CL.
%
%
\section{Complementary analysis: WMAP in the same region}
WMAP is a NASA satellite dedicated to observe the anisotropies of the
CMB with high accuracy at five different frequencies between 23 and 94
GHz.  Scientific results of this mission have allowed us to have a
clearer image of the early universe, and to reduce the uncertainties
in several cosmological parameters. Data products of this mission can
be found on the web\footnote{\tt http://lambda.gsfc.nasa.gov/}.
\subsection{The WMAP data}
We have analysed WMAP data with the same goodness-of-fit { and the
Minkowski functionals tests already} used on Archeops data. The main
purpose of this analysis is to compare Archeops results with a
different experiment to discriminate among systematics, foreground
emissions and intrinsic CMB non-Gaussian features. It is clear that
the WMAP frequencies complement very well the Archeops ones. A
detailed analysis of the possible WMAP non-Gaussianities with this
goodness-of-fit method deserves another work.

The maps that we have analysed have been produced from the 1-year and
3-year WMAP foreground cleaned maps for the differencing assemblies
corresponding to the cosmological frequencies 40, 60 and 90 GHz.  The
main properties of these maps are described in detail in
\citet{wmapbennet_1yr,wmapproc_3yr} respectively.

Specifically we have used the ``combined map'' as described in
\citet{wmapbennet_1yr}, \citep[see also][]{vielva_spot}. The
WMAP CMB simulations which are used in the analysis are also combined
simulations, that is, CMB signal simulations were produced for each
channel and then combined in the same manner than for the data.

According to \citet{wmapbennet_1yr} WMAP noise is highly uncorrelated,
that is, the noise from a given pixel $i$ is independent of the noise
from another pixel $j$. The noise combined simulations are produced
from the ``combined variance map'' as it is shown in
e.g. \citet{vielva_spot}.

\begin{figure}[t]
\center
\epsfig{file=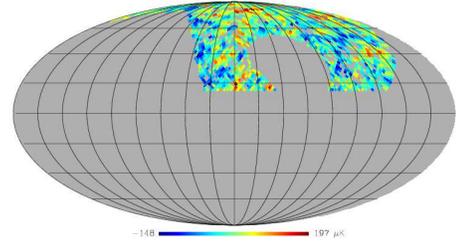,height=3.8cm,width=6.0cm}
\caption{WCM3 Data at HEALPix resolution $N_{side}=32$ 
(it corresponds to a pixel size of $\approx 1.8$ degrees).
This map is centered on Galactic longitude
$l$ = 180 degrees. The pixels contaminated by Galactic and extragalactic 
emission are covered with the mask described in the text. Grid lines are
spaced by 20 degrees.
}
\label{wmapskycoverage}
\end{figure}

We have analysed both combined maps, 1-year and 3-year (hereafter WCM1
and WCM3). The WMAP mask considered for both analyses was the 3-year
Kp0 one because it is the most conservative one for WCM3 and also
contains the 1-year Kp0 mask. See \citet{wmapproc_3yr} for details
about new masks and \citet{wmap_mask} for original masks.  The actual
mask we have used is the 3-year WMAP Kp0 degraded to our resolution
times the Archeops mask\footnote{ For comparison, we have also
repeated the goodness-of-fit analysis on Archeops data using this
combined mask, finding similar results to those obtained in section 5
using the Archeops mask.}. Its number of pixels is 1648. In figure
\ref{wmapskycoverage} WCM3 data is plotted using this mask.
\subsection{Gaussianity test on WMAP data}
\begin{figure*}[ht]
  \center
  \includegraphics[height=3.8cm,width=6.0cm]{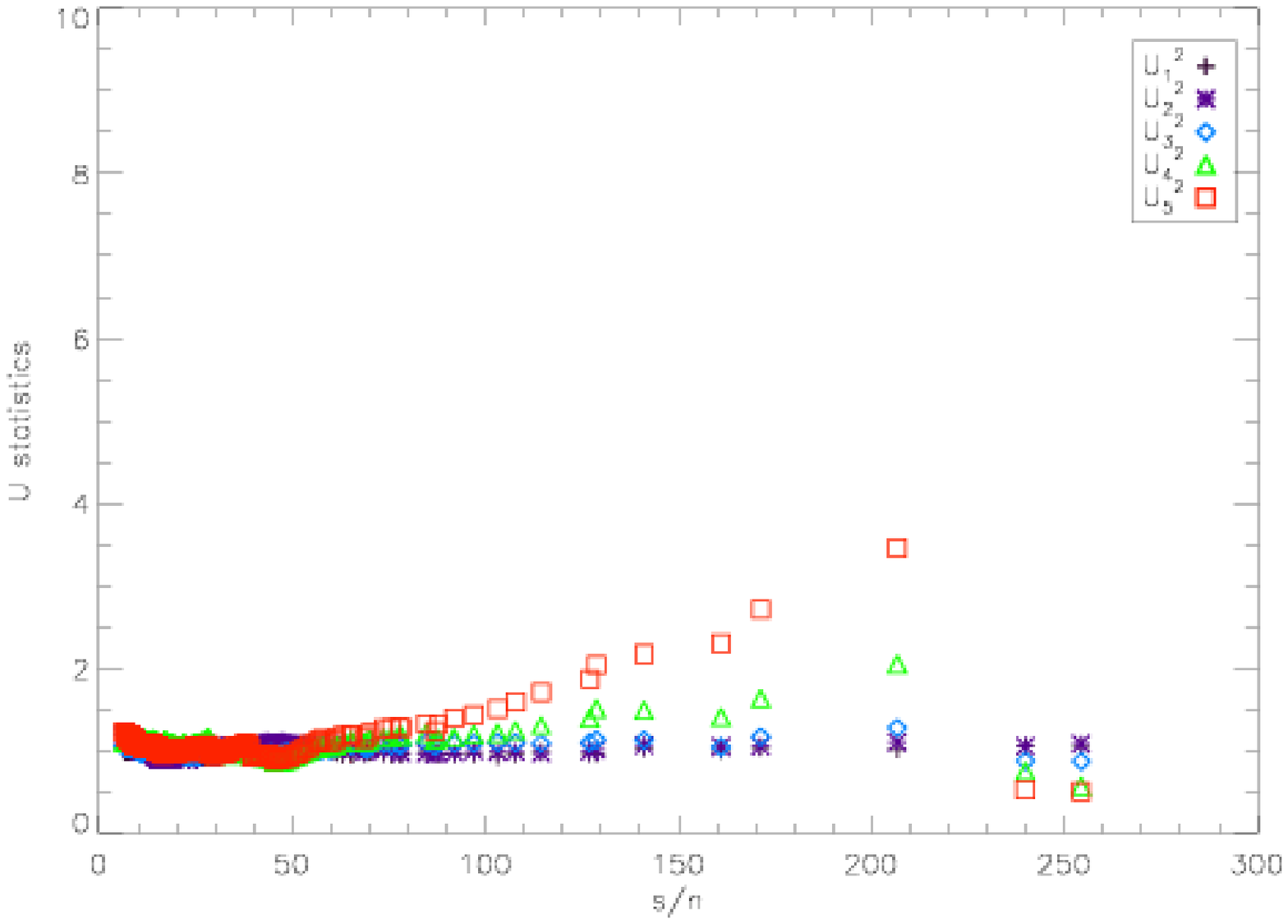}
  \includegraphics[height=3.8cm,width=6.0cm]{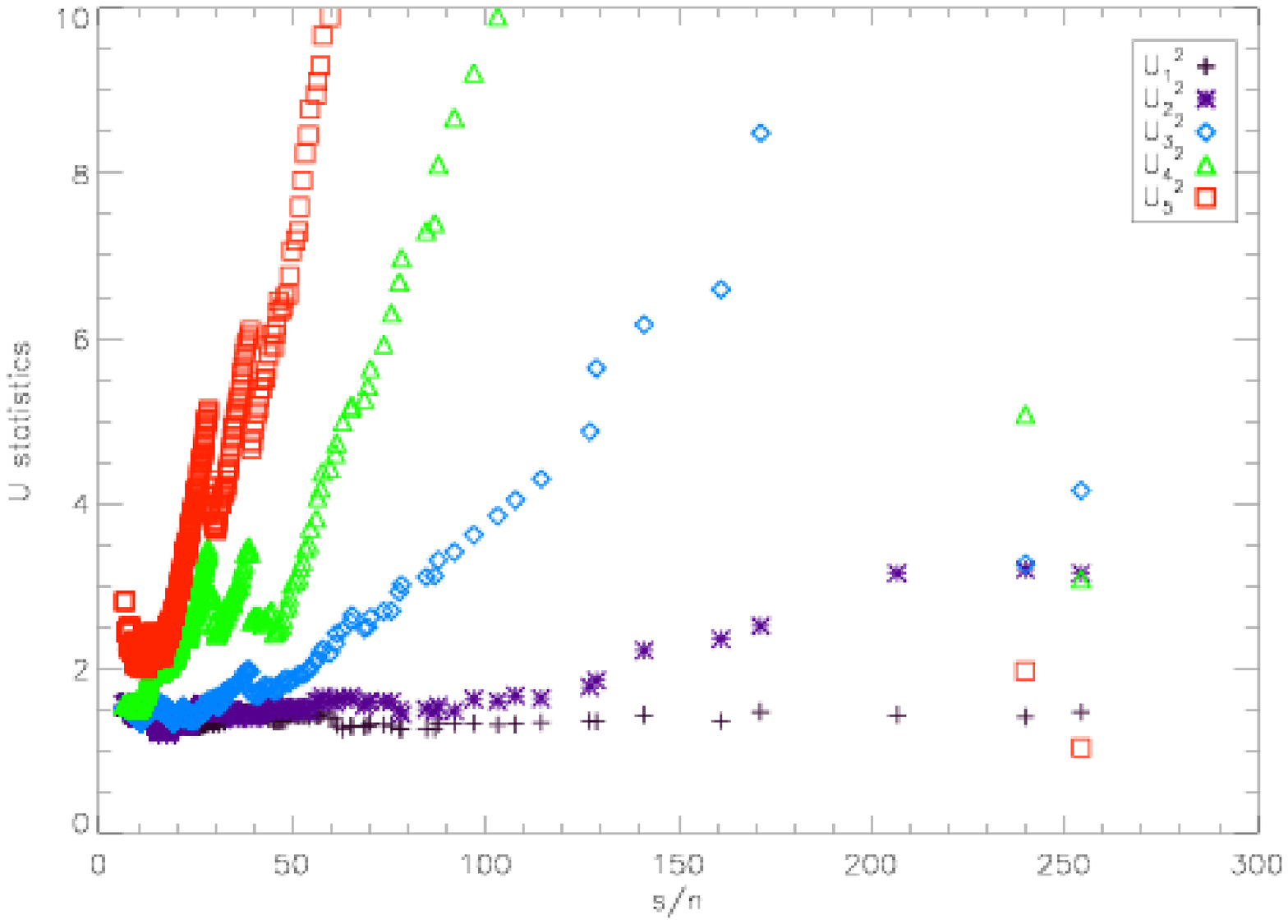}
  \caption{{\it \textbf{From left to right,}} mean and dispersion of
    $U_i^2$ statistics (where $i$ goes from 1 to 5) for different
    signal-to-noise cuts, corresponding to $10^3$ signal plus noise
    WCM3 simulations.
    \label{1st3rdwmap_simus}}
\end{figure*}
\begin{figure*}[th]
  \begin{center}
    \includegraphics[height=3.8cm,width=6.0cm]{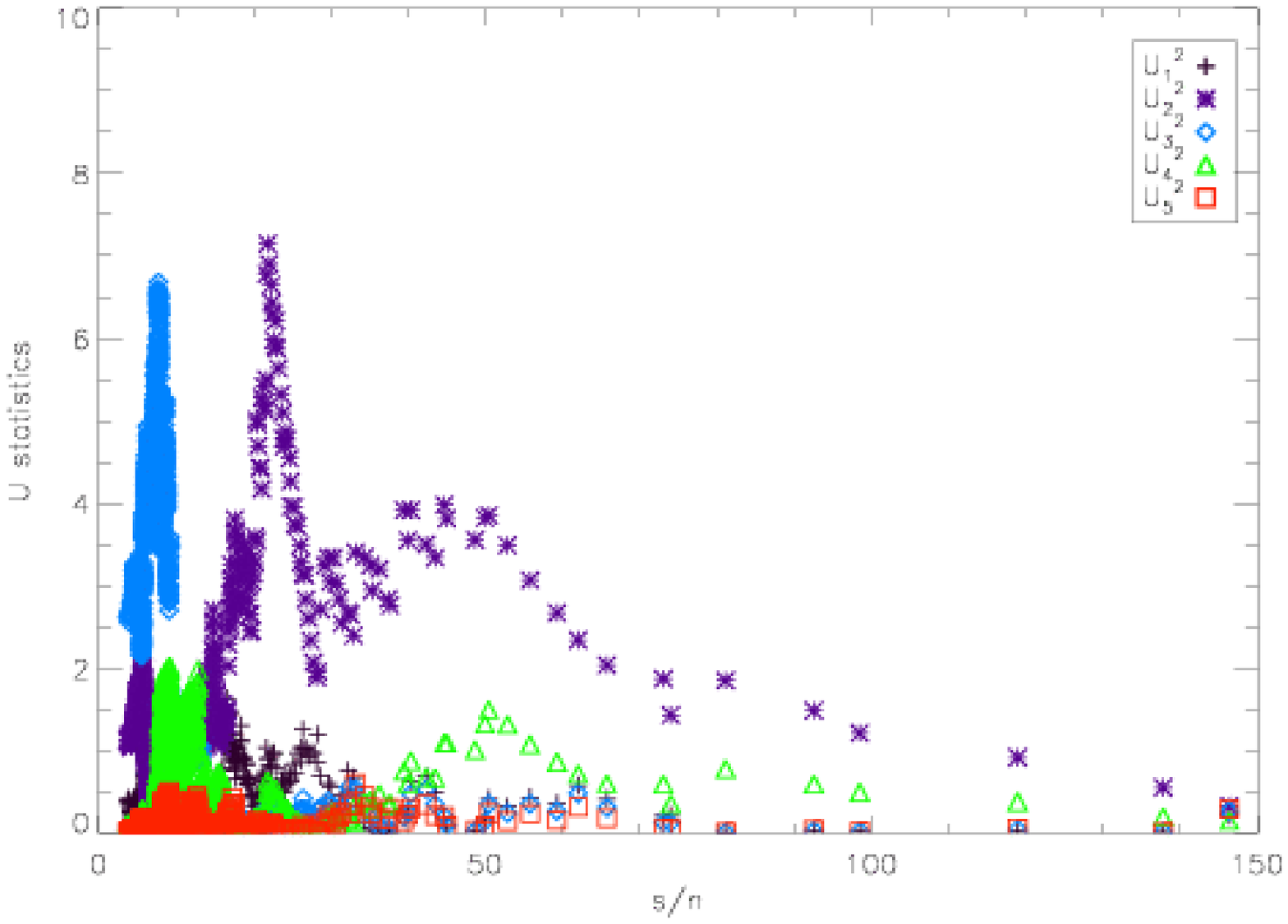}
    \includegraphics[height=3.8cm,width=6.0cm]{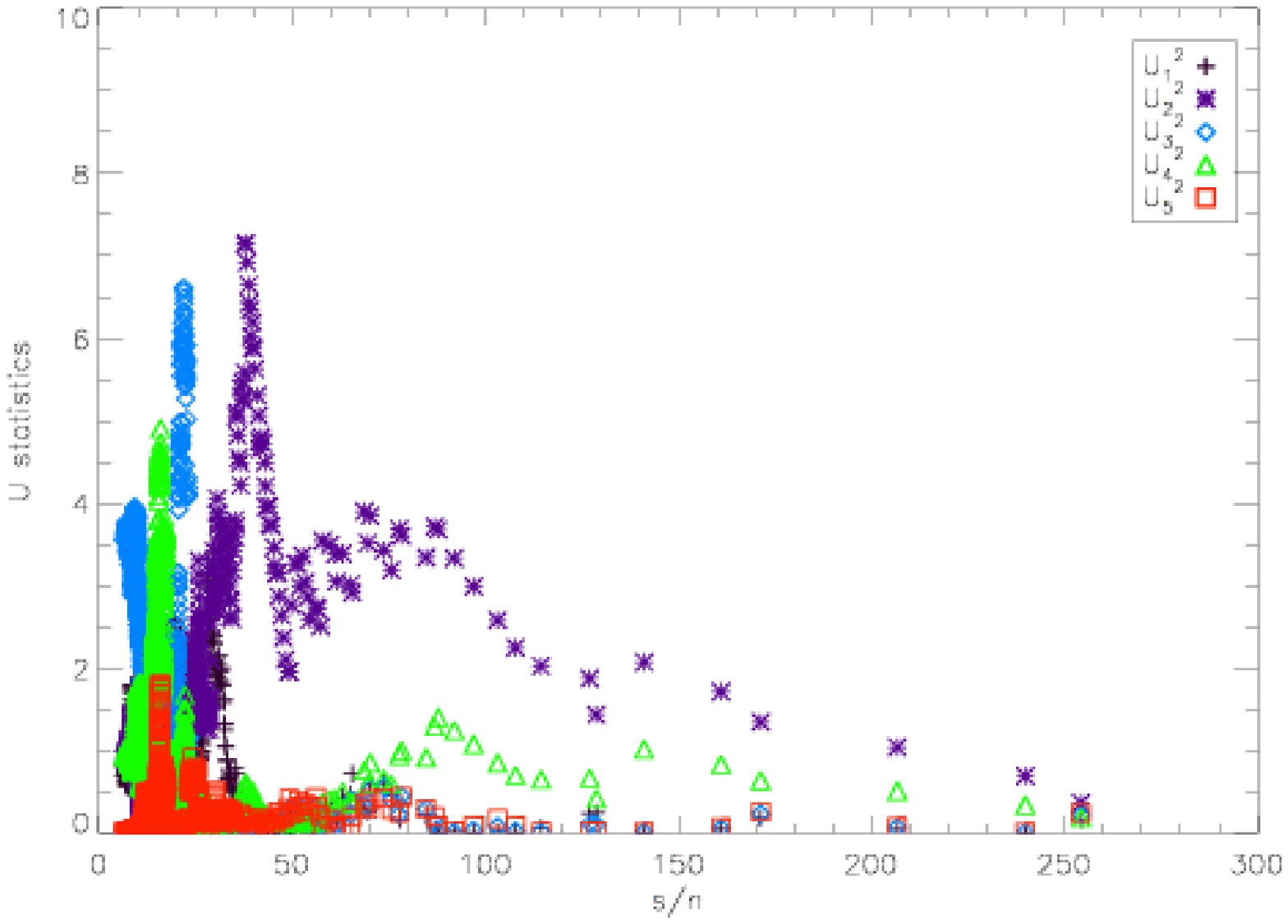}
    \caption{{\it \textbf{From left to right,}} $U_i^2$ statistics for
WCM1 and WCM3 presented for different signal-to-noise cuts.
\label{1st3rdwmap}}
  \end{center}
\end{figure*}
In order to perform the { R\&BT} test on WCM1 and WCM3 maps we
followed the same steps than for the Archeops analysis. We calculated
their corresponding $S$ and $N$ matrices for the 1648 pixels available
after applying the combined Archeops-WMAP mask.

We assume the best fit model of the 3-year WMAP data for both
analysis, WCM1 and WCM3. At the resolution with which we are
dealing, 1.8 degrees, the power spectra of the 1-year and 3-year data
are very approximately the same. This assumption implies that the $S$
matrix is the same for both releases. The $S$ matrix is computed from
$1.2 \times 10^5$ Gaussian simulations following equation
\ref{corrmatrix}.  Each simulation was produced in the same 90 dual
core processors mentioned before, and took an average CPU time of 360
s and an average RAM memory of 0.4 GB.

As commented above, WMAP noise is highly uncorrelated and therefore we
can assume that the noise matrices are diagonal.  This means that the
correlation element corresponding to pixels $i$ and $j$ is
$N_{ij}=\sigma^2_i*\delta_{ij}$, where $\sigma^2_i$ is the combined
noise of pixel $i$.  Noise matrices for WCM1 and WCM3 must be
constructed with their corresponding noise variances which differ by
an approximate factor of 3.

Two additional sets of $10^3$ Gaussian signal plus noise simulations
(corresponding to WCM1 and WCM3 maps) were performed for the
calibration of the matrices. In figure \ref{1st3rdwmap_simus}, we
present the mean and the dispersion of the $U_i^2$ statistics at
different signal-to-noise cuts for the WCM3 case.  Note that the
numerical range for the possible signal-to-noise cuts $(s/n)_c$ is
wider than for the Archeops case, because WCM3 noise is smaller than
the Archeops one at this resolution. $(s/n)_c$ range for WCM1 is
approximately the same than that of WCM3 reduced by a factor $\sqrt3$.
The mean and the dispersion for WCM1 simulations are similar to
those obtained for WCM3.
\begin{table}[th]
  \center
  \caption{Mean and dispersion of $U_i^2$ statistics from $10^3$ WCM1
    simulations for $(s/n)_c=3.64$. \label{table_stat_wmap}}
  \begin{tabular}{c|ccccc|c}
    \hline \hline
    ... & $U_1^2$ & $U_2^2$ & $U_3^2$ & $U_4^2$ & $U_5^2$ &  $\chi_1^2$ \\ 
    \hline
    $\mu$ & 1.09 & 1.15 & 1.02 & 1.09 & 1.02 &  1.00 \\
    $\sigma$ & 1.56 & 1.50 & 1.47 & 1.71 & 2.02 &  1.41 \\
    \hline
    \hline
  \end{tabular}
  \caption{Mean and dispersion of $U_i^2$ statistics from $10^3$ WCM3
    simulations for $(s/n)_c=6.33$. \label{table_stat_wmap3}}
  \begin{tabular}{c|ccccc|c}
    \hline \hline
    ... & $U_1^2$ & $U_2^2$ & $U_3^2$ & $U_4^2$ & $U_5^2$ &  $\chi_1^2$ \\
    \hline
    $\mu$ & 1.00 & 1.18 & 1.04 & 1.10 & 1.22 &  1.00 \\
    $\sigma$ & 1.42 & 1.56 & 1.51 & 1.56 & 2.81 &  1.41 \\
    \hline
    \hline
  \end{tabular}
\end{table}
It can be seen that mean values of $U_i^2$ statistics are close to one
for almost all signal-to-noise cuts and all the computed statistics,
but the dispersion becomes higher than square root of two for
high signal-to-noise cuts and for statistics with high order moments,
like $U_5^2$ and higher order statistics.

As for the Archeops case, these high values are explained by the small
errors present in the computed correlation matrices plus small
numerical errors in the diagonalization of these matrices, which are
amplified through the high order moments.  In table
\ref{table_stat_wmap} we present the mean and the dispersion of
$U_i^2$ statistics for $10^3$ WCM1 simulations with noise for all the
eigenmodes ($s/n \geq 3.64$).  Note how the dispersion is increasing
with the order of the statistics.  In table \ref{table_stat_wmap3} the
same quantities are presented for $10^3$ WCM3, obtained also from all
the eigenmodes ($s/n\ge 6.33$).  The effect is the same for the high
order moment statistics.  The results for the $U_i^2$ statistics for
WCM1 and WCM3 data maps are presented in figure \ref{1st3rdwmap}.  As
can be seen, all $U_i^2$ values satisfy $U_i^2 \le 7.15$.  The upper
limit 7.15 corresponds to a upper tail probability of 0.7\% for the
theoretical distribution.  In order to confirm or rule out a possible
non-Gaussian detection, this result should be studied more carefully.
\begin{figure}[t]
\center
\epsfig{file=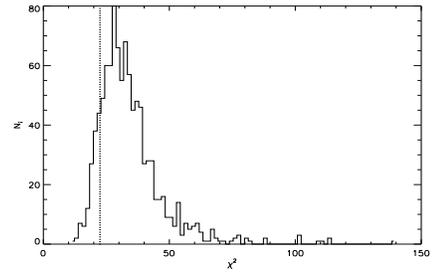,height=3.8cm,width=6.0cm}
\caption{Distribution of the $ \chi^2$ values from the
  Minkonwski Gaussian test for WCM3 data.  Vertical line shows the data
  results. Their cumulative probability is 12.0\%.}
\label{mkinkowskchiwcm3}
\end{figure}
\begin{table}[th]
  \center
  \caption{WCM1 $U_i^2$ statistics for $(s/n)_c=21.81$, and their corresponding
    upper tail probabilities.\label{table_probwcm1}}
  \begin{tabular}{c|ccccc}
    \hline \hline
    ... & $U_1^2$ & $U_2^2$ & $U_3^2$ & $U_4^2$ & $U_5^2$  \\
    \hline
    WCM1 & 0.90 & 7.15 & 0.32 & 0.63 & 0.09 \\
    \textbf{Prob.} & 0.37 & 0.01 & 0.52 & 0.35 & 0.67 \\
    \hline
    \hline
  \end{tabular}
  \caption{WCM3 $U_i^2$ statistics for $(s/n)_c=37.92$, and their corresponding
    upper tail probabilities.\label{table_probwcm3}}
  \begin{tabular}{c|ccccc}
    \hline \hline
    ... & $U_1^2$ & $U_2^2$ & $U_3^2$ & $U_4^2$ & $U_5^2$ \\
    \hline
    WCM3 & 0.13 & 7.15 & 0.00 & 0.61 & 0.01 \\
    \textbf{Prob.} & 0.73 & 0.01 & 0.95 & 0.36 & 0.88 \\
    \hline
    \hline
  \end{tabular}
  \end{table}
First of all, we have that for both WCM1 and WCM3 $U_2^2$ is the only
statistic which reaches some sharp peaks above $6.6$ (which
corresponds to a upper tail probability for the theoretical
distribution of 1.0\%).  From the plots in figure \ref{1st3rdwmap},
$U_2^2$ reaches this peak at $(s/n)_c=21.81$ for WCM1 and
$(s/n)_c=37.92$ for WCM3.  We estimated the upper tail probability for
the $U_i^2$ statistics of data at the mentioned signal-to-noise cut by
performing $10^3$ Gaussian simulations.  These results are presented
in tables \ref{table_probwcm1} and \ref{table_probwcm3}. As we can see
for the $U_2^2$ statistic, we have that this probability is 1.0\% and
0.7\% for WCM1 and WCM3 respectively, very similar to the theoretical
value.

This probability is obtained for the precise signal-to-noise cut where
$U_2^2$ reaches its maximum. Since the width of the maxima is much
smaller than the range of variation of the signal-to-noise
eigenvalues, it makes sense to ask for the significance of the
detection. Thus, from the simulations we computed the ``p-value'',
i.e.  the probability of finding a value of $U_2^2$ larger than 7.15
at any signal-to-noise cut, the maximum value reached by the
data. This probability is 18\% for WCM1 and 17\% for WCM3.

From the previous discussion, we conclude that the sharp peaks found
in the data are not significant. Also, well studied cases of
artificial CMB non-Gaussianities, like skewness or kurtosis produced
using the Edgeworth expansion (see \citet{edgeworth} for applications
of this expansion to the CMB non-Gaussianity analyses), usually show
deviations of the $U_i^2$ statistics in the form of a large plateau.
Besides, we would like to remark that at the signal-to-noise cuts
where the maxima are found there are less than one hundred $\{ y_i \}$
numbers to compute the $U_i^2$ statistics (around 70), and the test
works correctly only asymptotically ($n >> 1$).

WCM3 data were also analysed with the Minkowski functionals as in the
Archeops case (that is, using 11 thresholds between $ -2.5\sigma$ and
$ 2.5\sigma$ and the three functionals). The histogram corresponding
to the $ \chi^2$ values for 1000 Gaussian simulations and the value
for WCM3 data are presented in the figure \ref{mkinkowskchiwcm3}. As
we can see the WCM3 data are again compatible with Gaussianity.

Finally, we performed an analysis on simulations with $f_{nl}$
parameter as defined in equation \ref{nogchi}. The procedure was the
same as the one performed for Archeops case.  As discussed in
section 5.2, we only use the Minkowski functionals for the $
f_{nl}$ case. The $ \chi^2$ value for WCM3 data is minimum for $
f_{nl}$ = 100. Analysing Gaussian simulations, the constraints found
for $ f_{nl}$ are: $ f_{nl} = 100_{~-200}^{~+200}$ at 68\% CL,
$ f_{nl} = 100_{~-300}^{~+400}$ at 90\% CL, $ f_{nl} =
100_{~-400}^{~+500}$ at 95\% CL.
 These limits are compatible with those obtained from Archeops since
the tighter constraints found for WCM3 can be explained by the
significantly smaller noise in that experiment. In particular, if we
analyse simulated Archeops data with noise normalized to the same
amplitude as that of WCM3 we find similar limits for $ f_{nl}$.
\section{Conclusions}
The expected behaviour of the $U_i^2$ statistics as a $\chi_1^2$
distribution has been confirmed for the order index interval 
$1 \le i \le 4$ with ``realistic'' simulations assuming Gaussian CMB
anisotropies.  For higher moments, $i>4$, the mean of the distribution
is $\mu \simeq 1$ but the variance is $\sigma \gsim 2$. This is
because of the propagation of errors through higher order moments
which in practice complicates the use of high order $U_i^2$ in
our analysis.

From the analysis of both kind of Archeops maps, coaddition and
Mirage, we have found that both are compatible with Gaussianity. Only
the $U_2^2$ statistic for coaddition map is close to 8.0 for low
$(s/n)_c$.  Although in principle the probability that $U_2^2$ takes
values bigger than 8.0 for a given signal-to-noise cut in the Gaussian
hypothesis is very low (see table \ref{tabcoadd}), the corresponding
``p-value'' for having $U_2^2$ larger than 8.0 at any
signal-to-noise cut is 0.1482. This is not negligible and thus this
detection is not significant.  Moreover this effect does not appear in
the Mirage map, and therefore should be assigned to issues related to
the map-making process.

The analysis with the Minkowski functionals on the Mirage map
also returns compatibility with Gaussianity.

Our analysis also implies constraints on the amount of contamination
that can be present at 143 GHz. Using as template for dust and
atmosphere the Archeops map at 353 GHz, we limit the possible
contamination to be lower than { 7.8\%} at 90\% CL { using $
U_2$ statistic. A similar limit is obtained with the Minkowski
functionals.}

We have also compared the Archeops results with the WMAP 1 and 3-year
data in the same region of the sky. For both sets of data a sharp peak
in $U_2^2$ has been found at specific signal-to-noise cuts. Although
the probability of finding such a peak at a given signal-to-noise cut
is very small, the ``p-value'' obtained when different cuts are
allowed is appreciable. Therefore we can conclude that the WMAP data,
when the same region than Archeops is considered, are also consistent
with Gaussianity. { The same conclusion is reached when the data
are analysed with the Minkowski functionals.}

Finally, we have established a constraint in the value of the
non-linear coupling parameter $f_{nl}$.  Analysing Archeops data, {
we found that $ f_{nl} = 200_{~-600}^{~+900}$ at 90\% CL, and $
f_{nl} = 200_{~-800}^{~+1100}$ at 95\% CL}.  When the same analysis
was done with WMC3 data using Archeops-WMAP combined mask, { we
found $ f_{nl} = 100_{~-300}^{~+400}$ at 90\% CL, $f_{nl} =
100_{~-400}^{~+500}$ at 95\% CL. These limits are similar to the ones
expected for an Archeops-like experiment with a noise amplitude
similar to that of WCM3.}
\begin{acknowledgements}
The authors kindly thank the Archeops Collaboration for the
possibility of using Archeops data.  We are specially grateful for the
access to the computational resources of the IFCA Computing Group.  AC
thanks the Spanish Ministerio de Educaci\'on y Ciencia (MEC) for a
pre-doctoral FPI fellowship, P. Vielva for his help and comments on
the simulation of WMAP, and A. M. Aliaga for useful discussions on the
Smooth tests of goodness-of-fit.

We acknowledge partial financial support from the Spanish MEC project
ESP2004-07067-C03-01 and the joint project CSIC-CNRS, with reference
2004FR009. We acknowledge the use of LAMBDA. Support for
it is provided by the NASA Office of Space Science. We have used the
CAMB code \citep{camb} for our analysis. The CAMB code is derived from
CMBFAST \citep{cmbfast}.
The HEALPix package was used throughout the data analysis \citep{healpix}.

\end{acknowledgements}

\end{document}